\documentclass[a4paper,12pt]{article}
\usepackage[paper=a4paper,left=25mm,right=25mm,top=25mm,bottom=25mm]{geometry} 

 \usepackage{setspace}
\usepackage{authblk}

\usepackage{fullpage}
\usepackage{parskip}
\usepackage{titlesec}
\usepackage[section]{placeins}
\usepackage[dvipsnames]{xcolor}

\usepackage{breakcites}
\usepackage{lineno}
\usepackage{hyphenat}
\usepackage[all]{nowidow}

\PassOptionsToPackage{hyphens}{url}
\usepackage[colorlinks = true]{hyperref} 

\hypersetup{
  linkcolor  = RedViolet,
  citecolor = Blue!90!black,
  urlcolor   = Aquamarine!80!black,
  colorlinks = true
}

\usepackage{etoolbox}
\usepackage{natbib}
\renewenvironment{abstract}
  {{\bfseries\noindent{\abstractname}\par\nobreak}\footnotesize}
  {\bigskip}
\titlespacing{\section}{0pt}{*3}{*1}
\titlespacing{\subsection}{0pt}{*2}{*0.5}
\titlespacing{\subsubsection}{0pt}{*1.5}{0pt}

\usepackage{graphicx}
\usepackage[space]{grffile}
\usepackage{latexsym}
\usepackage{textcomp}
\usepackage{longtable}
\usepackage{tabulary}
\usepackage{booktabs,array,multirow}
\usepackage{amsfonts,amsmath,amssymb}
\usepackage{subcaption}
\providecommand\citet{\cite}
\providecommand\citep{\cite}
\providecommand\citealt{\cite}
\newif\iflatexml\latexmlfalse

\AtBeginDocument{\DeclareGraphicsExtensions{.pdf,.PDF,.eps,.EPS,.png,.PNG,.tif,.TIF,.jpg,.JPG,.jpeg,.JPEG}}

\usepackage[utf8]{inputenc}
\usepackage[english]{babel}
\usepackage{float}

\newcommand{\blockForest}{BlockForest}
\newcommand{\Clinicalonly}{Clinical Only}
\newcommand{\CoxBoost}{CoxBoost}
\newcommand{\CoxBoostfavoring}{CoxBoost Favoring}
\newcommand{\ipflasso}{Ipflasso}
\newcommand{\grridge}{Grridge}
\newcommand{\glmboost}{Glmboost}
\newcommand{\KaplanMeier}{Kaplan-Meier}
\newcommand{\Lasso}{Lasso}
\newcommand{\ranger}{Ranger}
\newcommand{\rfsrc}{Rfsrc}
\newcommand{\prioritylasso}{Prioritylasso}
\newcommand{\prioritylassofavoring}{Prioritylasso Favoring}
\newcommand{\linebreakcust}{\\} 
\usepackage{orcidlink}
\newcommand{\beginsupplement}{
        \setcounter{table}{0}
        \renewcommand{\thetable}{S\arabic{table}}
        \setcounter{figure}{0}
        \renewcommand{\thefigure}{S\arabic{figure}}
     }

\begin{document}
\title{Over-optimism in benchmark studies and the multiplicity of design and analysis options when interpreting their results}

\def\correspondingauthor{\footnote{Corresponding author, e-mail: \href{mailto:cniessl@ibe.med.uni-muenchen.de}{cniessl@ibe.med.uni-muenchen.de}, Institute for Medical Information Processing, Biometry and Epidemiology, Ludwig Maximilians University Munich, Marchioninistr. 15, D-81377, Munich, Germany.}}
\author[1]{Christina Nie{\ss}l \correspondingauthor{} \orcidlink{0000-0003-2425-7858}}
\author[2]{Moritz Herrmann \orcidlink{0000-0002-4893-5812}}
\author[1]{Chiara Wiedemann}
\author[2]{Giuseppe Casalicchio \orcidlink{0000-0001-5324-5966}}
\author[1]{Anne-Laure Boulesteix \orcidlink{0000-0002-2729-0947}}
\affil[1]{Institute for Medical Information Processing, Biometry and Epidemiology, Ludwig Maximilians University Munich (Germany)}
\affil[2]{Department of Statistics, Ludwig Maximilians University Munich (Germany)}
\vspace{-1em}
  \date{\today}
\maketitle

\begin{abstract}

In recent years, the need for neutral benchmark studies that focus on the comparison of methods from computational sciences has been increasingly recognised by the scientific community. While general advice on the design and analysis of neutral benchmark studies can be found in recent literature, certain amounts of flexibility always exist. This includes the choice of data sets and performance measures, the handling of missing performance values and the way the performance values are aggregated over the data sets. As a consequence of this flexibility, researchers may be concerned about how their choices affect the results or, in the worst case, may be tempted to engage in questionable research practices (e.g. the selective reporting of results or the post-hoc modification of design or analysis components) to fit their expectations or hopes. To raise awareness for this issue, we use an example benchmark study to illustrate how variable benchmark results can be when all possible combinations of a range of design and analysis options are considered. We then demonstrate how the impact of each choice on the results can be assessed using multidimensional unfolding. In conclusion, based on previous literature and on our illustrative example, we claim that the multiplicity of design and analysis options combined with questionable research practices lead to biased interpretations of benchmark results and to over-optimistic conclusions. This issue should be considered by computational researchers when designing and analysing their benchmark studies and by the scientific community in general in an effort towards more reliable benchmark results.

\end{abstract}

\sloppy

\section{Introduction and related work}
With the constant development of new methods in computational sciences (e.g. machine learning and bioinformatics), it is becoming increasingly difficult for data analysts to keep pace with scientific progress and to select the most appropriate method for their data and research question out of the many existing approaches.  
This problem is addressed by benchmark studies, which systematically analyse and compare the performance of several methods in different conditions using simulated or real data sets.\\
In many cases, benchmark studies are performed as part of a paper introducing a new method, usually with the intention to demonstrate the superiority of the new method over existing ones. Accordingly, they can be considered as biased in favour of the newly proposed method and should be seen as an informal method comparison rather than a real benchmark study  \citep{Norel2011,Boulesteix2013,Buchka2021}. In contrast, so called \textit{neutral} benchmark studies are defined as benchmark studies that focus on the comparison itself and are ideally performed by reasonably neutral authors, i.e. authors  who (1) are 
equally experienced with all considered methods and (2) design and analyse the study in a rational way \citep{Boulesteix2017}.
These characteristics make neutral benchmark studies essentially  unbiased.  Therefore, recommendations resulting from such studies are  especially relevant both for method users and developers \citep{Boulesteix2018}. \\
Regarding the appropriate design and analysis of benchmark studies, the available literature ranges from general guidelines \citep{Weber2019, Boulesteix2015} and statistical frameworks \citep[][all with focus on supervised learning]{Demsar2006, Hothorn2005,  Eugster2012,  Boulesteix2015a}, to recommendations for context-specific benchmarks \citep[e.g.][]{Mangul2019,Bokulich2020,Zimmermann2020, Kreutz2019}. 
However, for many issues relevant in practice (e.g. the selection of data sets and performance measures), no concrete guidance or methodology can be found. This means that researchers are usually faced with a high amount of flexibility when conducting their benchmark study.\\
As a consequence, researchers who are aware of these issues, although making well-considered design and analysis choices prior to conducting the benchmark study, might be concerned about how their choices affect the results. 
On the other hand, the high amount of flexibility could tempt less aware researchers to engage in questionable research practices (see \citealp{John2012}, in the context of applied research) when conducting their benchmark study. This includes
the selective reporting of results (e.g. reporting the results of only one performance measure although performance was originally assessed by two measures)
and the modification of specific design and/or analysis components of the benchmark study after seeing the results (e.g. using performance measures other than those originally selected). 
Of course, these practices are not questionable on their own. For example, it is fine to use an alternative performance measure if the current one does not produce meaningful results as long as the change of performance measure is adequately justified and documented.
However, practices such as the selective reporting of results or the post-hoc modification of benchmark components do become questionable if they are applied to fit the researchers' expectations or hopes. For example, researchers might seek an ``exciting'' result (e.g. a clear-cut result suggesting a univocal winner as opposed to vague tendencies) or have a specific presumption in mind that they want to be confirmed by the results (e.g. the superiority of a certain method or class of methods that they are more familiar with or that has performed well in previous benchmark studies).\linebreakcust
The problem with such research practices is that they are likely to produce over-optimistic results, i.e. results with an optimistic bias towards the researchers' expectations and hopes. 
While we are convinced that very few researchers have the actual intention to cheat \citep{Ioannidis2014}, it should not be understated that ``even an honest person is a master of self-deception'' \citep{Nuzzo2015}, meaning that every researcher is at risk of 
engaging in questionable research practices. Moreover, the non-neutrality that leads to such practices in the first place is difficult to avoid completely and is likely to arise in a subconscious manner even in studies intended as neutral.  Note also that the actual neutrality of neutral benchmark studies can only be checked to a certain extent. For example, one may review the authors' publication lists to identify the methods they are most familiar with, but this gives only a partial picture of someone's (non-)\-neutrality. \linebreakcust
In application fields of statistics (e.g. medicine and psychology), the multiplicity of analysis strategies and the associated risk of over-optimistic results are well-known issues \citep{Ioannidis2005, Simmons2011, Hoffmann2021} and terms such as ``p-hacking'' or ``fishing expeditions'' have been discussed by many \citep{Head2015, Wagenmakers2012}.
However, in methodological research including benchmark studies, this topic is covered rather sparsely.
Existing literature on the risk and prevention of over-optimism in benchmark studies is either limited to general considerations in benchmarking guidelines \citep{Weber2019, Boulesteix2017} or to benchmark studies that are performed as part of a paper introducing a new method \citep{Norel2011,Boulesteix2015}, which can be transferred to neutral benchmark studies only to a limited extent.
Similarly, the scarce literature that {\it empirically} investigates the effects of over-optimism in benchmark studies in a quantitative manner is either also devoted to the bias affecting evaluations of a newly proposed method to other existing methods \citep{Jelizarow2010,Buchka2021}, or focusing on the selection of data sets \citep{MacIa2013, Yousefi2010}.  \\
In this paper, we illustrate and discuss the multiplicity of options regarding the design and analysis of neutral benchmark studies based on real data sets and examine its effect on the results. Note that although we focus on neutral benchmark studies based on real data, our results are also relevant to benchmarks comparing new to existing methods and, to some extent, benchmarks based on simulated data. 
We will empirically address the multiplicity of options and its effects in a twofold approach. In the first step, in order to raise awareness of the multiplicity of possible results and the over-optimism that may arise from questionable research practices, we use the results of a recently published benchmark study to illustrate how variable the resulting method rankings are when different options for design and analysis are considered. 
In the second step, we propose a framework based on multidimensional unfolding \citep{Borg2005} that enables researchers to assess the impact of each choice on the method rankings. 
More precisely, the framework allows to analyse when and how using alternative options for a specific choice affects the results and can thus be an effective strategy to prevent biased interpretations and over-optimistic conclusions. \\
The exemplary study we will use throughout the paper to illustrate our proposed framework and the multiplicity of possible options and results is a benchmark experiment by \cite{Herrmann2020} comparing the performance of 13 survival prediction methods based on 18 real so-called \lq\lq multi-omics'' data sets.
Note that our paper does not intend to question the results of this study. Instead, it should be seen as extended analysis of the benchmark study, which by assessing the multiplicity of results and examining the impact of each choice, makes the results of \cite{Herrmann2020} even more reliable and meaningful. \\
The remainder of this paper is structured as follows: we review and discuss a selection of design and analysis choices in the context of benchmark studies in Section~\ref{02_design_analysis_choice}, and describe the design of the study as well as the principle of multidimensional unfolding in Section~\ref{03_methods}. The results are presented in Section~\ref{04_results}, which is followed by a discussion in Section~\ref{05_discussion} and concluding remarks in Section~\ref{06_conclusion}.

\section{Examples of design and analysis choices in benchmark studies}\label{02_design_analysis_choice}
\subsection{Setting}
\begin{figure}[ht]
    \centering
    \includegraphics[width = 1\textwidth]{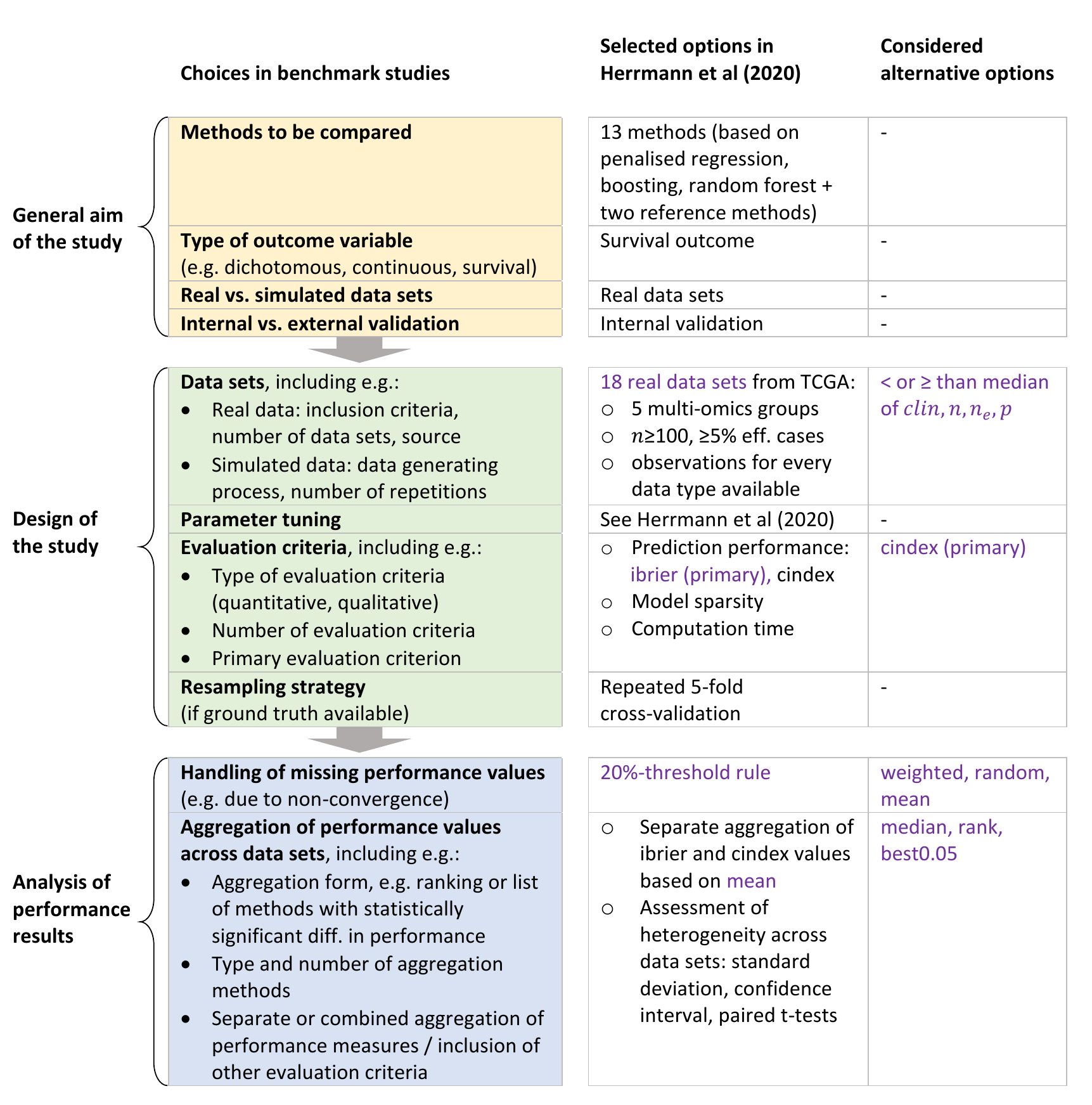}
    \caption{Examples of choices that researchers are usually faced with when conducting a benchmark study including options used in the example benchmark study by \protect\cite{Herrmann2020} (second column) and alternative options (third column). Options that are considered in our illustration are coloured in purple.} 
    \label{fig:choices}
\end{figure}

 In this section, we discuss some of the choices that researchers are faced with when conducting a benchmark study based on real data sets. In general, most choices that have to be made to conduct a benchmark study relate to (1) the general aim of the study, (2) the design of the study, or (3) the analysis of the performance results; see the left part of Figure~\ref{fig:choices}. Choices that belong to the first category are, for instance, the choice of methods to be compared or the type of outcome variable to be considered. However, in this paper, we focus on choices regarding the design of the study (i.e. how the aim of the study is addressed) and the analysis of performance results (i.e. how the $L\times M$ matrix of results generated by each considered performance measure is analysed, where $L$ and $M$ are the numbers of data sets and methods, respectively). It is important to note that these choices should ideally be made prior to conducting the benchmark study. However, we conjecture that they are in practice often made post-hoc, i.e. after seeing the results---which can amount to questionable research practices. 
 When reading a benchmark study, there is no way to check when the choices were made.\\
For each choice, we will give concrete examples of possible options that will later be analysed with regard to their effect on the results; see the right part of Figure~\ref{fig:choices}. For this purpose, we consider the benchmark study by \cite{Herrmann2020} mentioned above. The authors compare the performance of  $M=13$ survival prediction methods (here denoted as \textit{\blockForest}, \textit{\Clinicalonly}, \textit{\CoxBoost}, \textit{\CoxBoostfavoring}, \textit{\glmboost}, \textit{\grridge}, \textit{\ipflasso}, \textit{\KaplanMeier}, \textit{\Lasso}, \textit{\prioritylasso}, \textit{\prioritylassofavoring}, \textit{\ranger} and \textit{\rfsrc}) on $L=18$ real multi-omics data sets.  See the original paper  \citep{Herrmann2020} for details on the methods, the benchmark experiment and the results.
\\
\subsection{Design choices}
\subsubsection{Data sets}
The selection of data sets is an important design choice in every benchmark study, as the performances are usually highly variable  across data sets \citep{Weber2019,Novianti2015}. To make meaningful statements and prevent the study from being underpowered, it is recommended to consider an adequate number of data sets  \citep{Boulesteix2017}.
 Although there are suggestions on how to calculate the minimum required number  \citep{Boulesteix2015a}, it seems that the number of included data sets is usually based on practical criteria (such as availability or computational cost) rather than statistical considerations \citep{MacIa2013}. 
 Moreover, if the benchmark study aims at external validation, the number of data sets that can be included in the benchmark study is usually limited, as for many data sets there is often no comparable data set available that could be used for external validation.\\ 
 Concerning the type of data sets, researchers should include data sets that are representative for the domain of interest and diverse enough to make sure the methods can be evaluated under a wide range of conditions \citep{Gatto2016,Weber2019}.
 Corresponding inclusion criteria for the data sets should be defined before conducting the benchmark study  \citep{Boulesteix2017}. 
 However, the decision on how the inclusion criteria are defined lies with the researcher.
In many benchmark studies,  the exact search strategy or inclusion criteria are not reported transparently, suggesting that in these cases, there might be no clearly defined inclusion criteria at all. \\
In the benchmark study by \cite{Herrmann2020}, the authors selected all cancer data sets with five different multi-omics groups and more than 100 samples from the TCGA research network (\url{http://cancergenome.nih.gov}). Additionally, they excluded data sets that did not have observations for every data type or less than 5\% effective cases (i.e. patients with event), resulting in a total of $L=18$ data sets.
 However, depending on their research interest, \cite{Herrmann2020} could have set additional constraints. For example, if the authors had been interested in the performance of the methods on data sets with a small number of effective cases, they could have adjusted the inclusion criteria accordingly (e.g. set $n_e < 30$). The other way around, one may decide to ignore data sets with a small number of events  (e.g. set $n_e \geq 30$) because it is questionable if it makes sense to fit models in this case at all.\\
 In this paper, we will address the multiplicity of possible options regarding the selection of data sets and its impact on the results by considering subgroups of the original $L=18$ data sets defined based on some of the data sets' characteristics. The considered characteristics are the number of clinical variables ($clin$), the number of observations ($n$), the number of effective observations ($n_e$), and the number of variables ($p$). 
 For each data set characteristic, we will only consider data sets that are smaller ($<$) or greater or equal ($\geq$) than the median value of the respective data set characteristic over the 18 considered data sets. This results in eight groups with 8 to 10 data sets.

\subsubsection{Quantitative performance measure}\label{02_performance}
Another important aspect of benchmarking is the choice of evaluation criteria, which usually includes both quantitative performance measures and other measures such as runtime or qualitative features such as user-friendliness. Although all these evaluation criteria are important, we will focus on quantitative performance measures in this paper. \\
The choice of performance measure is usually context-specific, i.e. it depends on the  type of methods and data addressed in the benchmark study, as well as on the aspects of performance that are considered the most important by the researcher \citep{Weber2019,Morris2019}.
It is also often a non-trivial choice. For some tasks such as classification, researchers are spoilt for choice considering the variety of measures they can choose from (e.g. accuracy, sensitivity/specificity, area under the curve or F1-Score), which makes decisions difficult \citep{Mangul2019, Robinson2019}. 
In contrast, for more complex situations they might have to design their own performance measures, which can also be challenging  \citep{Weber2019}.
To provide a more complete picture of the methods' behaviour and avoid over-optimism, it can be useful to consider more than one performance measure \citep{Norel2011}. However, there is no way to objectively determine the adequate number of performance measures as this is highly context dependent.\\
In the benchmark study by \cite{Herrmann2020}, the primary performance measure is the integrated Brier score (\citealp{Graf1999}; denoted as \textit{ibrier}). Additionally, they consider Uno's C-index (\citealp{Uno2011}; denoted as \textit{cindex}).
The authors justify their decision to use the ibrier as primary measure by the fact that cindex only measures the discriminatory power and is not a strictly proper scoring rule \citep{Blanche2019}, while the ibrier additionally measures calibration. However, they argue that if the main interest lies in {\it ranking} patients according to their risk, then the cindex would also be a valid measure. Furthermore, they reason that it makes sense to include the cindex for the purpose of comparability with other studies, since it is a widely used performance measure.
Accordingly, depending on which aspect of performance they would have considered more important, \cite{Herrmann2020} could have also used the cindex as primary performance measure or only selected one of the two performance measures.
In this paper, we will thus compare the results of ibrier and cindex.
\subsection{Analysis choices}
\subsubsection{Handling of missing performance values}
Because of non-convergence or other computational issues, methods sometimes fail to output a result for a specific data set. In the context of resampling procedures such as cross-validation or bootstrapping, the consequence is that performance values may be missing for all or part of the resampling iterations for some data sets. This problem seems to be common especially in benchmarks of larger scale \citep{Bischl2013}.
While there is at least some literature devoted to the selection of data sets and performance measures, the issue of missing performance values in some combinations of data sets and methods is almost completely ignored. 
Many authors of benchmark studies do not report how they handled missing performance values, and there is to our knowledge no corresponding guidance available.\\
 \cite{Bischl2013} mention several possible ad-hoc options that could be applied if the missing values occur only on a subset of resampling iterations, namely that missing values could be imputed by the worst possible value or by the mean of the remaining performance values obtained for this combination of data set and method---although both options are not ideal in their opinion. Another ad-hoc option they actually use for their benchmark study is a mixed strategy, where the imputed value is sampled from an estimated normal distribution of the remaining values if the method fails in less than 20\% of the resampling iterations. If the method fails in more than 20\% of the resampling iterations, the worst possible value is used for imputation.
\cite{Herrmann2020}, who use cross-validation as resampling procedure and also face the problem of failing iterations,
use a similar 20\%-threshold rule as \cite{Bischl2013}. 
However, instead of sampling from a normal distribution, they use the mean performance value of the remaining iterations and instead of the worst possible value, they assign values of the performance measures corresponding to random prediction (i.e. 0.25 for ibrier and 0.5 for cindex).\\
Since there seems to be no common agreement on how to handle missing values in this context, other sensible options would also be justifiable. For example, missing values could be imputed by a formula that weights the mean performance value and the random performance value used by \cite{Herrmann2020} according to the proportion of missing values, thus avoiding the choice of an arbitrary threshold. For the ibrier, where 0 corresponds to the best possible value and 0.25 to random prediction,
the imputed value for the considered combination of data set and method could be defined as
\begin{align}\label{eq: imp_method}
    x_{impute} = 0.25- (0.25 - \frac{\sum_{i\in \mathcal{I}}x_i}{|\mathcal{I}|})_+ \cdot(1-r),
\end{align}
where $\mathcal{I}$ is the set of indices of the non-failed iterations,  $x_i$ is the ibrier value for iteration $i \in \mathcal{I}$ and $r$ is the proportion of missing values. For two methods with the same mean value for non-failed iterations, the method with more missing values obtains a worse performance value.  Moreover, the imputed value is equal to 0.25 if a method has 100\% failures for a data set, or a mean value greater or equal than 0.25 (which makes sense since fluctuations above the value 0.25 corresponding to random prediction are not relevant). Another advantage of this weighted imputation procedure is that it reduces to the mean when the proportion of missing values $r$ tends to 0---as intuitively expected.   The corresponding formula for the cindex can be found in the supplementary material. \\
In this paper, we will consider four imputation methods that can be used to handle the issue of missing performance values:  the 20\%-threshold rule used by \cite{Herrmann2020}, the weighted method in Eq.~\eqref{eq: imp_method},  imputation using values that correspond to random prediction, and imputation using the average of the non-failed iterations.

\subsubsection{Aggregation of performance values across data sets}
Although it is common to analyse the methods' individual performances across data sets (e.g. using  graphical tools), most benchmark studies ultimately aggregate the performance values over the data sets to generate an overall method evaluation, for example in the form of a ranking or a list of methods that show statistically significant differences in performance. While there is much literature addressing statistical testing procedures in benchmark experiments based on a single data set  \citep{Dietterich1998,Hothorn2005} or several data sets \citep{Demsar2006,Eisinga2017}, there seems to be no consensus on how to generate an overall method {\it ranking} from several data sets, which we will focus on in this section.\\
For example, the performance values can be aggregated using standard summary measures such as the mean, median, minimum, maximum or standard deviation \citep{Mersmann2015}. 
Since the distribution of performance values can be considerably skewed, some authors advise against using the mean or median as aggregation method. 
Instead, they recommend assigning ranks to the methods for each data set such that the best method in the corresponding data set obtains rank~1 and the worst method rank~$M$, where $M$ is the number of considered methods  \citep{Demsar2006, Hornik2007}.
The resulting ranks are then usually aggregated using the mean \citep[e.g.][]{Kibekbaev2016, Verenich2019} or, less often, the median \citep[e.g.][]{Orzechowski2018}.\\
Other possible aggregation methods include counting the number of times a method performs best, often divided by the number of data sets to obtain a value between 0 and 1 \citep[e.g.][]{DeCnudde2020,Fernandez-Delgado2014, Wu2020}.
Some of these authors suggest to not only consider the best performing method for each data set but also the set of methods performing similarly to the best method. Accordingly, \cite{Fernandez-Delgado2014} consider the number of data sets in which a method achieves 95\% or more of the maximum accuracy (i.e. the accuracy achieved by the best performing method in that data set) divided by the total number of data sets. In the same vein, \cite{Wu2020} estimate the probability of achieving good performance as the number of data sets for which the method is among the top three methods divided by the total number of data sets.\\
Note that all aggregation methods presented so far are based on 
point estimates of the methods' performances. 
Although less frequently used in practice, it is also possible to 
generate method rankings based on the results of 
statistical tests (i.e. pairwise comparisons indicating if method~$1$ performs significantly better than method~$2$) using consensus rankings \citep{Hornik2007}. \\
If more than one performance measure and/or other evaluation criteria (e.g. runtime) are considered, researchers also have to decide if rankings arising from multiple criteria should be combined in some form \citep[e.g.][]{Eugster2012} or should be considered separately, as suggested by \cite{Weber2019}. Specifically, \cite{Weber2019} recommend to identify a set of consistently high performing methods based on the individual rankings and then highlight the different strengths of each method.  \linebreakcust
\cite{Herrmann2020} aggregate the performance values based on ibrier and cindex using the mean and consider each ranking separately. To assess the heterogeneity of performances across data sets, they also calculate the resulting standard deviations and confidence intervals and perform paired t-tests. 
In our illustration, we will consider four aggregation methods that can be used to generate method rankings: mean \citep[as used by][]{Herrmann2020}, median, mean rank and number of times a method performs best. If two methods obtain the same rank according to the number of times they perform best, they are additionally ranked by the number of times their performance lies within the 5\% environment of the best performing method. This applies if  $\frac{|\overline{x}_{m} - \overline{x}_{best}|}{\overline{x}_{best}} < 0.05$, where $\overline{x}_m$ denotes the performance (cindex or ibrier) of method $m$ and $\overline{x}_{best}$ the performance of the best performing method in the corresponding data set. We denote this aggregation method (i.e. counting the number of times a method performs best and the number of times it lies within the 5\% environment as secondary ranking method) as \textit{best0.05}. 
Note that since we only evaluate the results of one performance measure at a time (ibrier or cindex), we are not considering different options for combining rankings that result from more than one performance measure.

\section{Methods}\label{03_methods}
\subsection{Design of the study}
To illustrate the variability of benchmark results with respect to design and analysis choices, we use the benchmark results from \cite{Herrmann2020} and systematically examine different combinations of design and analysis options. Specifically, we consider all combinations of options regarding the choice of data sets (9 options), performance measure (2 options), imputation method (4 options), and aggregation method (4 options) described in Section~\ref{02_design_analysis_choice}  and Figure~\ref{fig:choices}. This results in $9\times2\times4\times4=288$ combinations. We then compare the 288 resulting rankings of the 13 survival prediction methods, where a rank of 1 corresponds to the best performing method and a rank of 13 to the worst performing method (average ranks are assigned in case of ties). 

\subsection{Multidimensional unfolding}
The impact of each choice on the method rankings is assessed using multidimensional unfolding \citep{Coombs1964, Borg2005}, which we will briefly introduce in the remainder of this section. Multidimensional unfolding is a technique that represents preference data as distances in a low-dimensional space.
It locates $K$ ideal points representing the subjects (in our case, $K=288$ combinations) and $M$ object points representing the objects (in our case, $M=13$ methods) such that the distances from each ideal point to the object points correspond to the observed preference values. The closer an object point lies to a subject's ideal point, the stronger the subject's preference for that object. Accordingly, the ideal point itself corresponds to maximal preference \citep{Borg2013}.\\
Multidimensional unfolding takes non-negative dissimilarities $\delta_{km}$ ($k= 1,\dots,K$; $m = 1,\dots, M$) as input, 
which are the preference values possibly converted in a way that small values correspond to high preferences. In our case, where the preference values are ranks, this is not necessary since a small rank already indicates  high preference. Moreover, the number of dimensions $dim$ must be specified, which we set to $dim=2$ as it is done in most applications of multidimensional unfolding. 
To find the coordinates for the points representing the $K$ subjects and $M$ objects, a loss function (\textit{stress}) is minimised.
It is defined as 
	 \begin{equation}\label{eq:stress}
	     \sigma^2(\hat{\boldsymbol{D}},\boldsymbol{Z}_1, \boldsymbol{Z}_2) = \sum_{k=1}^K \sum_{m=1}^M w_{km}(\hat{d}_{km} - d_{km}(\boldsymbol{Z}_1, \boldsymbol{Z}_2))^2,
	  \end{equation}
where $w_{km}$ denotes a non-negative a priori weight (which is set to $w_{km}= 1$ by default), and $\boldsymbol{Z}_1 \in \mathbb{R}^{K\times dim}$ and  $\boldsymbol{Z}_2 \in \mathbb{R}^{M\times dim}$ are the coordinates for the points representing the subjects and objects, respectively. Moreover, $d_{km}(\boldsymbol{Z}_1, \boldsymbol{Z}_2)$ denotes the fitted Euclidean distances
	\begin{equation} 
	        d_{km}(\boldsymbol{Z}_1, \boldsymbol{Z}_2) = \sqrt{\sum_{s=1}^{dim}(z_{1ks} - z_{2ms})^2}.
	 \end{equation} 
The matrix $\hat{\boldsymbol{D}} \in \mathbb R_{0}^{+ K\times M}$ contains the disparities $\hat{d}_{km} = f(\delta_{km})$, which are the optimally scaled dissimilarities. This means that the loss function in Eq.~\eqref{eq:stress} is not only minimised with respect to $\boldsymbol{Z}_1$ and $\boldsymbol{Z}_2$ but also with respect to a function $f(\cdot)$ that transforms the dissimilarities $\delta_{km}$ into disparities $\hat{d}_{km}$ (the function class depends on the assumed scale level). 
If the preference data is available in form of rankings, $f(\delta_{km})$ reflects a monotone step function that is found through monotonic regression on the dissimilarities. 
This type of multidimensional unfolding is referred to as ordinal or non-metric unfolding. 
To avoid degenerate solutions due to equal disparities which occur particularly often in non-metric unfolding, it is recommended to use a penalised version of the stress function in \eqref{eq:stress} that involves the coefficient of variation $v(\hat{\boldsymbol{D}})$. 
The penalised stress function is minimised through numerical optimisation using a strategy called SMACOF (Stress Majorization of a Complicated Function) and is implemented in an R  package of the same name \citep{rpackage_smacof}. For details on multidimensional unfolding and its implementation see \citeauthor{Mair2019} (forthcoming), \cite{Borg2005} and \cite{Busing2005}. \\

\section{Results}\label{04_results}
For full reproducibility, the entire analysis and the results presented in this section are publicly available in the GitHub repository \url{https://github.com/NiesslC/overoptimism_benchmark}.

\subsection{Overall variability and step-wise optimisation}\label{04_results1}
\begin{figure}[ht]
    \centering
    \includegraphics[width = 0.8\textwidth]{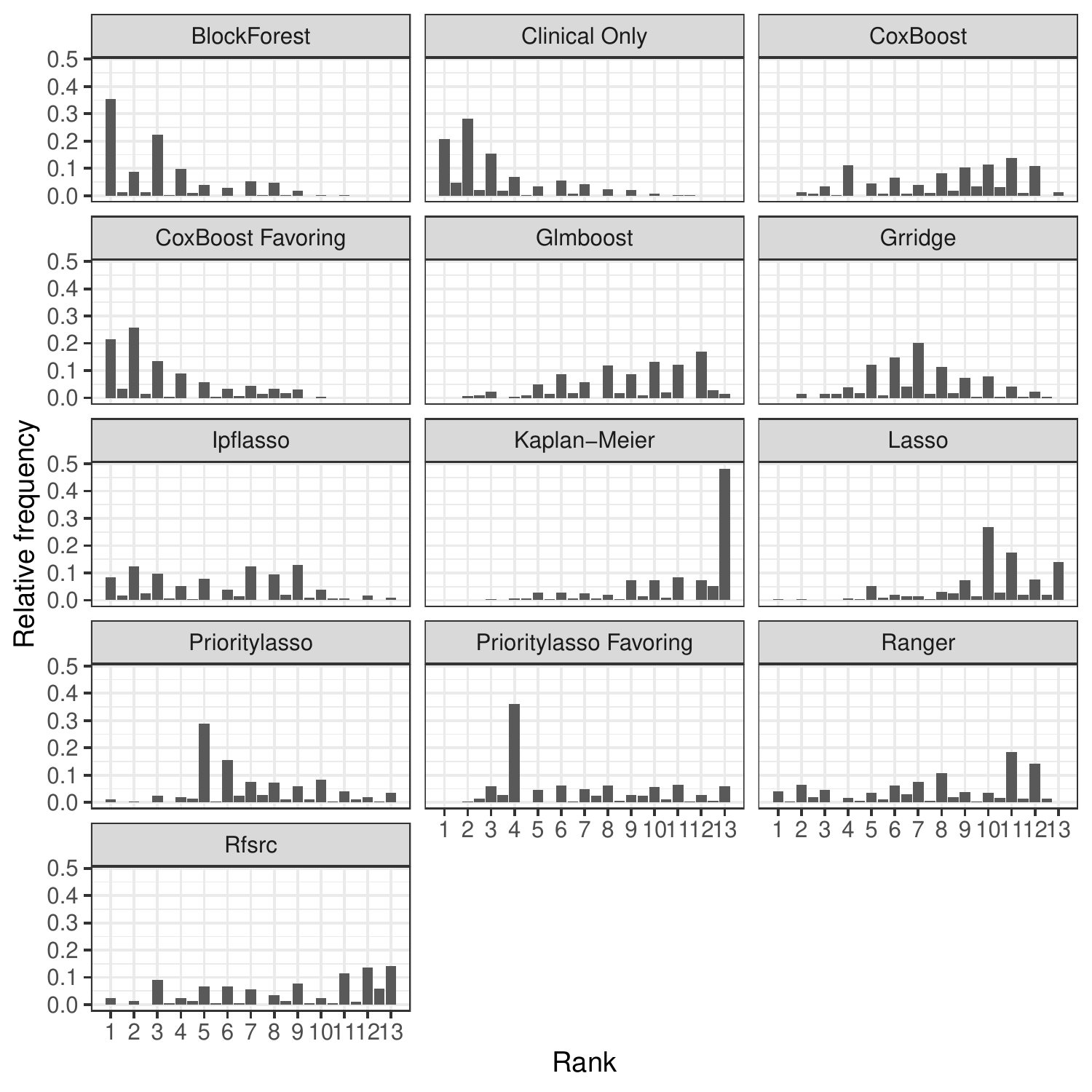}
    \caption{Rank distribution of 13 methods generated by 288 combinations of design and analysis options.}
    \label{fig:rankdistr}
\end{figure}

\begin{figure}[ht]
	\centering
	\includegraphics[trim= 0cm 0cm 0.2cm 0cm, clip, width=0.9\textwidth]{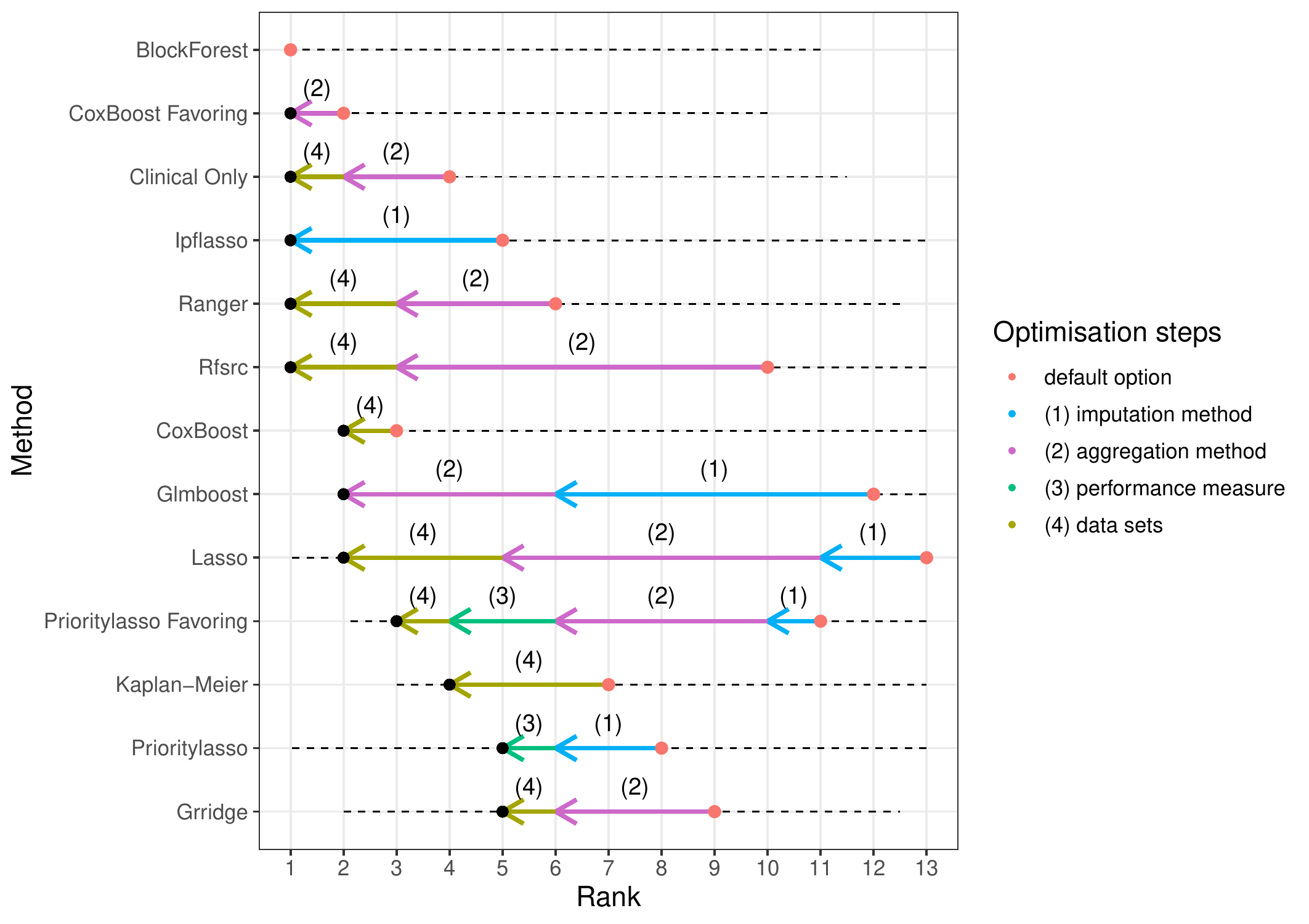}
	\caption{Step-wise optimisation of method ranks by (1) imputation method (blue), (2) aggregation method (pink), (3) performance measure (green), and (4) data sets (yellow). The dotted line corresponds to the smallest and highest possible ranks when all 288 combinations are considered. Missing steps indicate that they did not lead to an improved rank. Default options correspond to \protect\cite{Herrmann2020}.}
			\label{fig:stepwise imp_agg_eval_group} 
\end{figure}

As a first step, we compare the method rankings resulting from all 288 combinations of design and analysis options. 
Figure~\ref{fig:rankdistr} shows the corresponding rank distribution  for each method. Importantly, it reveals that any method can achieve almost any rank. On one hand, all methods but one achieve rank 1 (8 methods) or 2 (4 methods) for at least one combination. The exception is \KaplanMeier, which does not use any feature information  and can achieve ranks as small as 3. On the other hand, 10 methods are found to be the worst or one of the two worst methods (i.e. have rank 13 or 12.5, respectively) for at least one combination. The highest rank obtained by the remaining methods (\Clinicalonly, \blockForest, and \CoxBoostfavoring) ranges from 10 to 11.5. Figure~\ref{fig:rankdistr} also reveals that the ranks are distributed differently for each method. For example, while \Clinicalonly\ obtains rank 1 or 2 in approximately 50\% of the combinations, the ranks of \ranger\ are more  evenly distributed. 
\\
While considering all combinations of options provides valuable information on the overall variability of results, it is not a realistic scenario concerning over-optimism in the sense that no researcher conducting a benchmark study would try all possible combinations to obtain a favourable result (unless they are actively cheating, which we do not assume here). Therefore, we additionally illustrate how easy it is to modify the method rankings if the design and analysis options are selected in a step-wise optimisation process, which might represent a more realistic scenario. In our illustration, the step-wise optimisation for each method is performed as follows: In each step (i.e. for each choice), the option that yields the best rank for the considered method  (or the best performance value in case of equal ranks) is selected. If all options yield the same result, the default option is used. 
As default options, we use all 18 data sets, ibrier as primary performance measure,  the 20\%-threshold rule as imputation method, and the mean as aggregation method. This corresponds to the setting of \cite{Herrmann2020}.
Moreover, we assume that a favourable result is a small rank for a specific method. Note that this may not always be the case, e.g. if one expects a reference method such as \KaplanMeier\ to obtain a high rank or considers a group of several methods as target. \linebreakcust
Figure~\ref{fig:stepwise imp_agg_eval_group}  displays the optimisation process if the ranks are optimised in the order: (1) imputation method, (2) aggregation method, (3) performance measure, and (4) data sets.
It shows that for 8 of 13 methods, the best rank achieved by step-wise optimisation corresponds to the smallest possible rank for the corresponding method (i.e. the smallest rank that can be achieved when all 288 combinations are considered) and for another three methods, the step-wise optimisation achieves one rank higher than the smallest possible rank. Only two methods (\prioritylasso\ and \grridge) show a larger discrepancy between step-wise optimisation and considering all possible combinations. However, this is not too surprising considering the few cases and thus very specific combinations where they achieve small ranks (see Figure~\ref{fig:rankdistr}). 
If a step is missing in the optimisation process of a certain method, this indicates that the corresponding step did not improve the rank of that method. In fact, all methods except \Lasso\ and \prioritylassofavoring\ require no more than two optimisation steps. \\
Note that the results of the step-wise optimisation depend on the default options. For example, when cindex instead of ibrier is used as default option, the resulting ranks are higher (see Figure~\ref{fig:stepwise imp_agg_eval_group_cindex} in the supplementary material).  Moreover, the results depend on the order in which the ranks are optimised. The order shown in Figure~\ref{fig:stepwise imp_agg_eval_group} is realistic in the sense that researchers might find it more problematic to modify components of the benchmark study that are generally considered as important (i.e. performance measure or data sets) and thus only resort to them if the previous optimisation steps (i.e imputation method or aggregation method) do not yield a favourable result.
However, other orders in which the ranks are optimised would also be conceivable. For example, the selection of data sets could be optimised first since it offers many options and can be easily modified by eliminating specific data sets. In this case, the selection of data sets remains the only optimisation step for many methods since the subsequent steps do not lead to an improvement  (see Figure~\ref{fig:stepwise group_eval_imp_agg} in the supplementary material), which already indicates the large impact of data set selection,  discussed in more detail in the next section.
\subsection{Impact of individual design and analysis choices}\label{04_results2}
\begin{figure}[ht]
	\centering
	\subfloat[Performance measure]{\label{fig:unfolding_performance}
		\includegraphics[trim= 0.7cm 0cm 0.56cm 0cm, clip, width=0.48\textwidth]{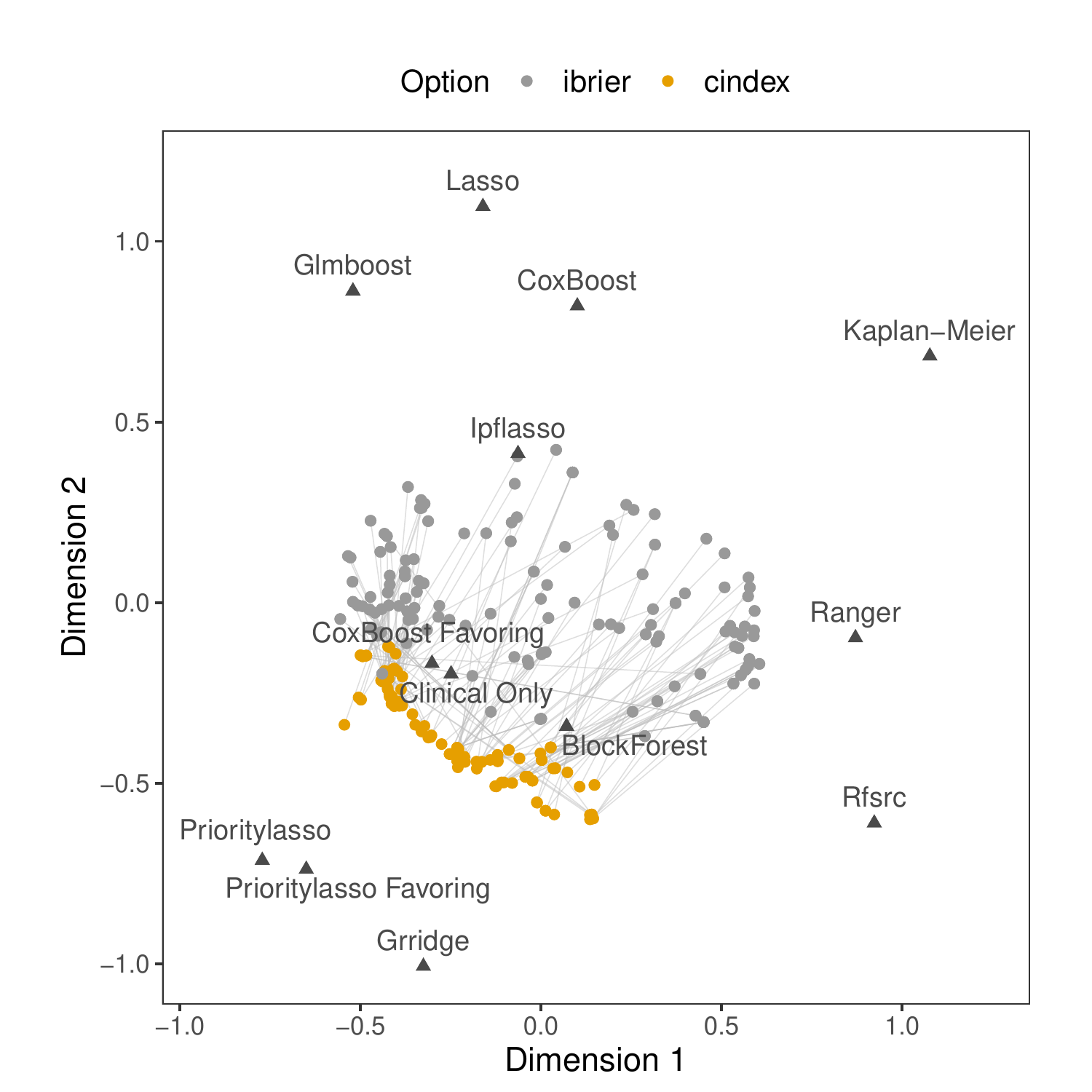}}
	\subfloat[Data sets]{\label{fig:unfolding_datasets}
		\includegraphics[trim= 0.7cm 0cm 0.56cm 0cm, clip, width=0.48\textwidth]{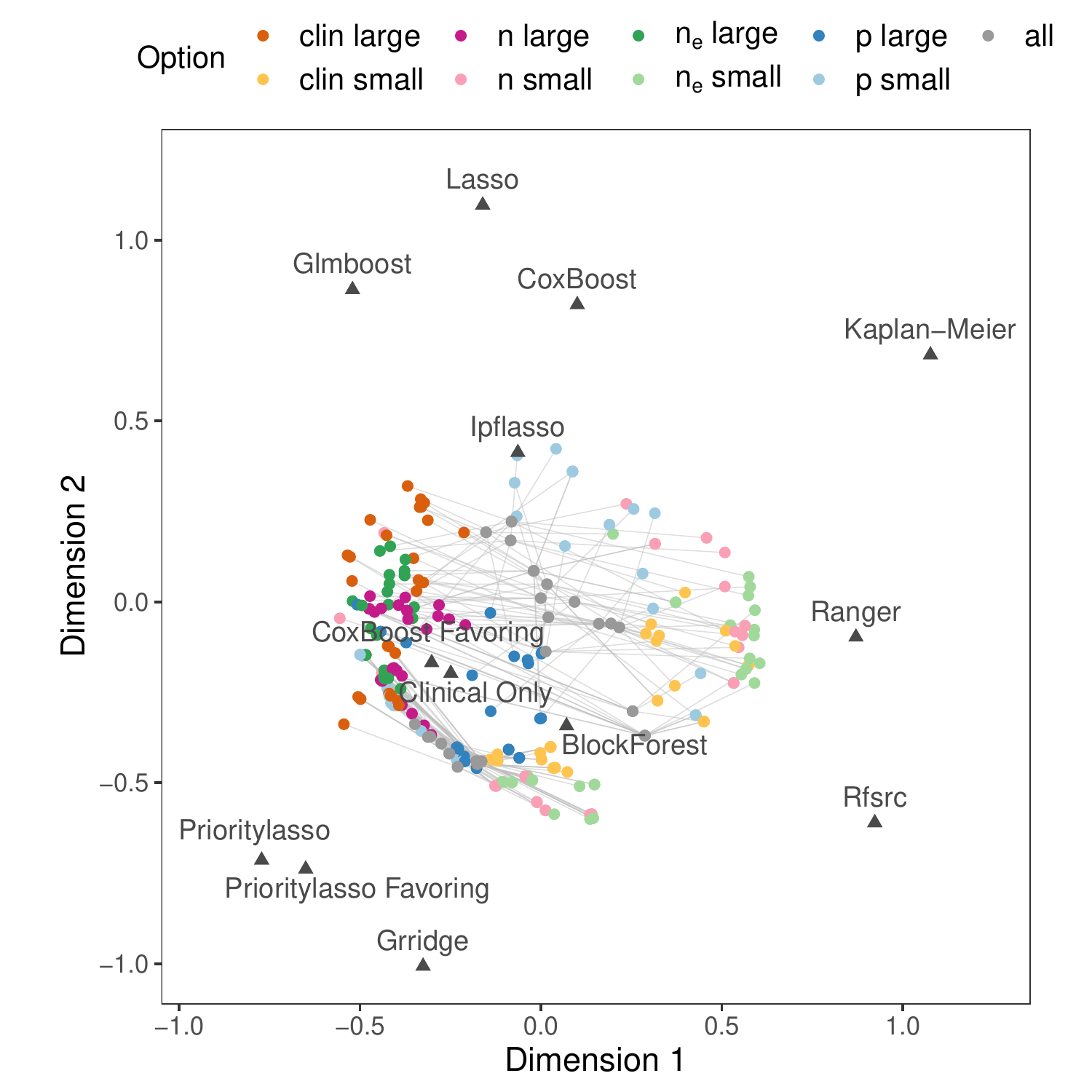}}\\
	\subfloat[Imputation method]{\label{fig:unfolding_imputation}
		\includegraphics[trim= 0.7cm 0cm 0.56cm 0cm, clip, width=0.48\textwidth]{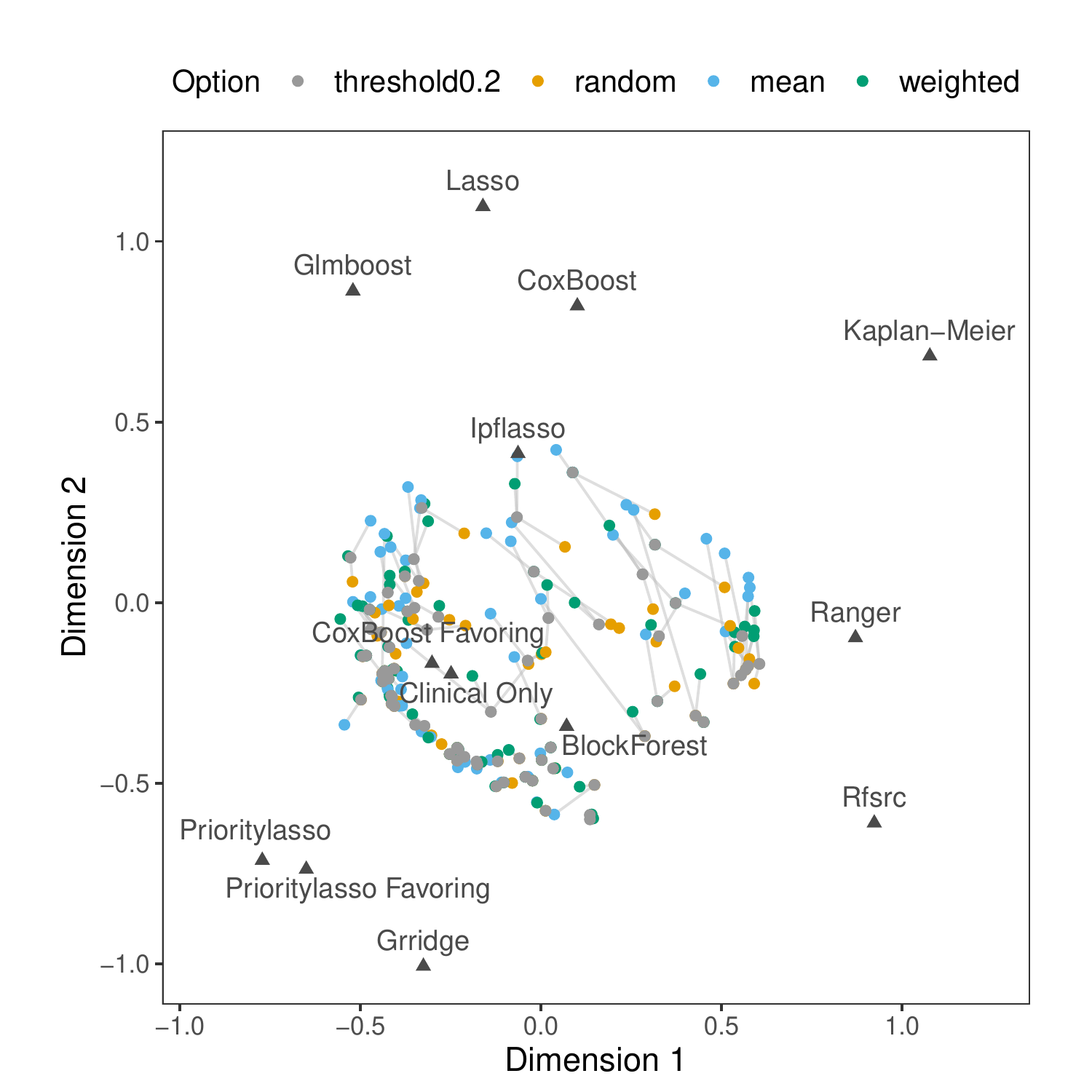}}
	\subfloat[Aggregation method]{\label{fig:unfolding_aggregation}
		\includegraphics[trim= 0.7cm 0cm 0.56cm 0cm, clip, width=0.48\textwidth]{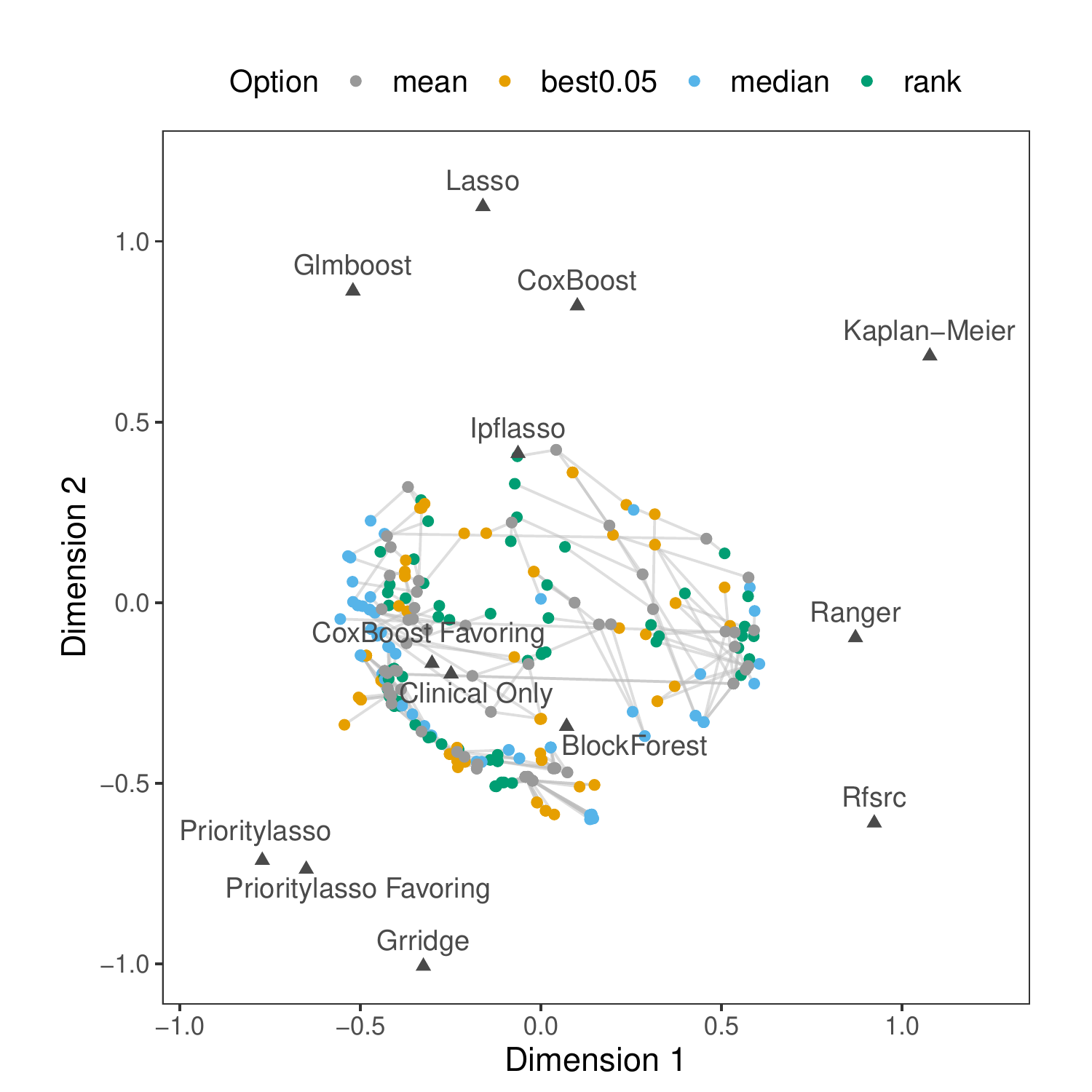}}
	\caption{Unfolding solution representing the rankings of 288 combinations of design and analysis options (\textit{ideal points}; circles) regarding 13 methods (triangles). For each choice, the ideal points are coloured according to the option that was used in the respective combination (default options corresponding to \protect\citealp{Herrmann2020} are grey). Each ideal point representing a default option is connected to the ideal points representing alternative options, given that the other three choices remain the same.}
	\label{fig:unfolding}
\end{figure}

\begin{figure}[ht]
    \centering 
    \includegraphics[width = 0.9\textwidth]{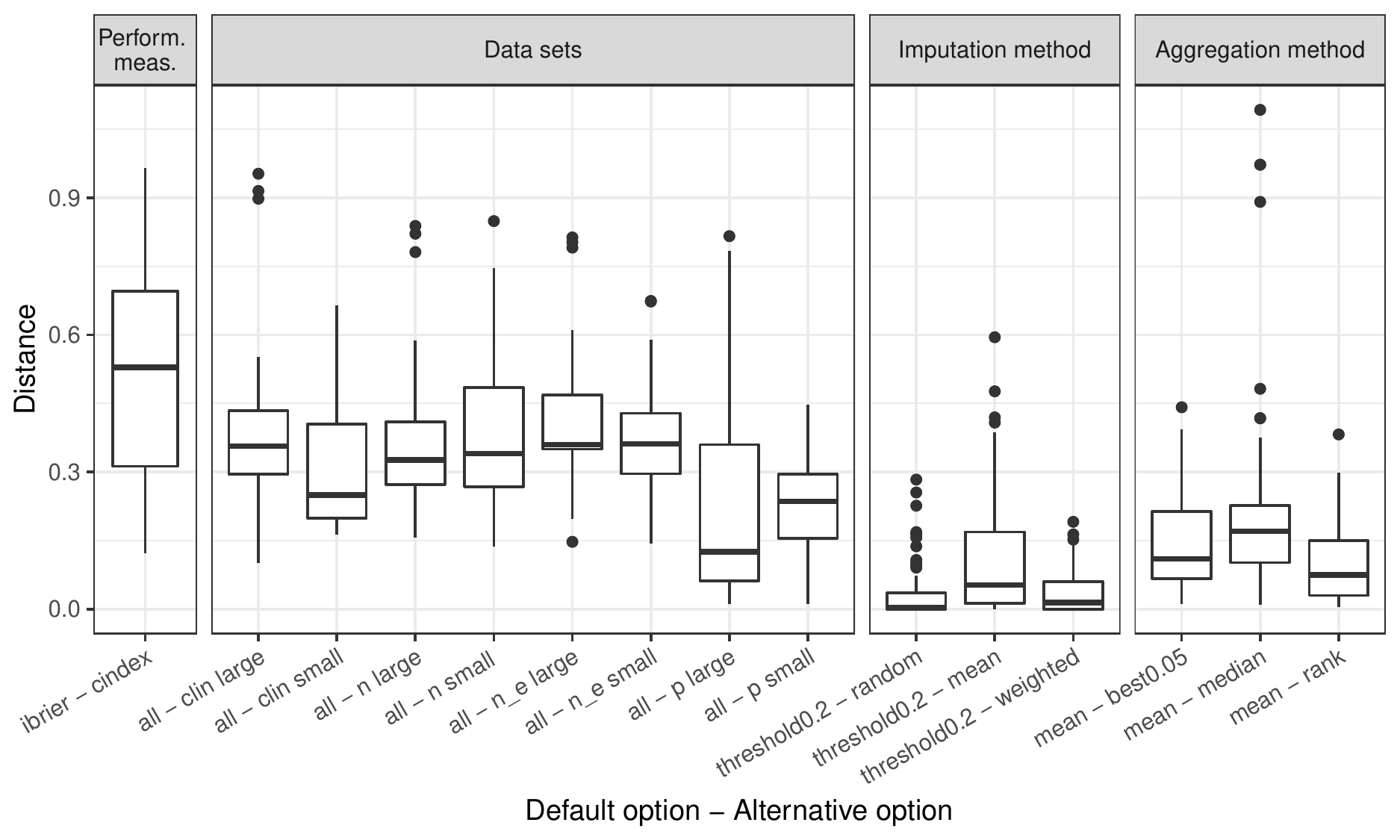}
    \caption{Distances between ideal points of combinations that represent default and alternative options of one specific choice (given that the other three choices remain the same),  derived from the unfolding solution in Figure~\ref{fig:unfolding}. The larger the distance, the larger the discrepancy between the two rankings generated by using the alternative option instead of the default option.}
    \label{fig:unfolding_dist}
\end{figure}

\begin{figure}[ht]
    \centering
    \includegraphics[width = 0.9\textwidth]{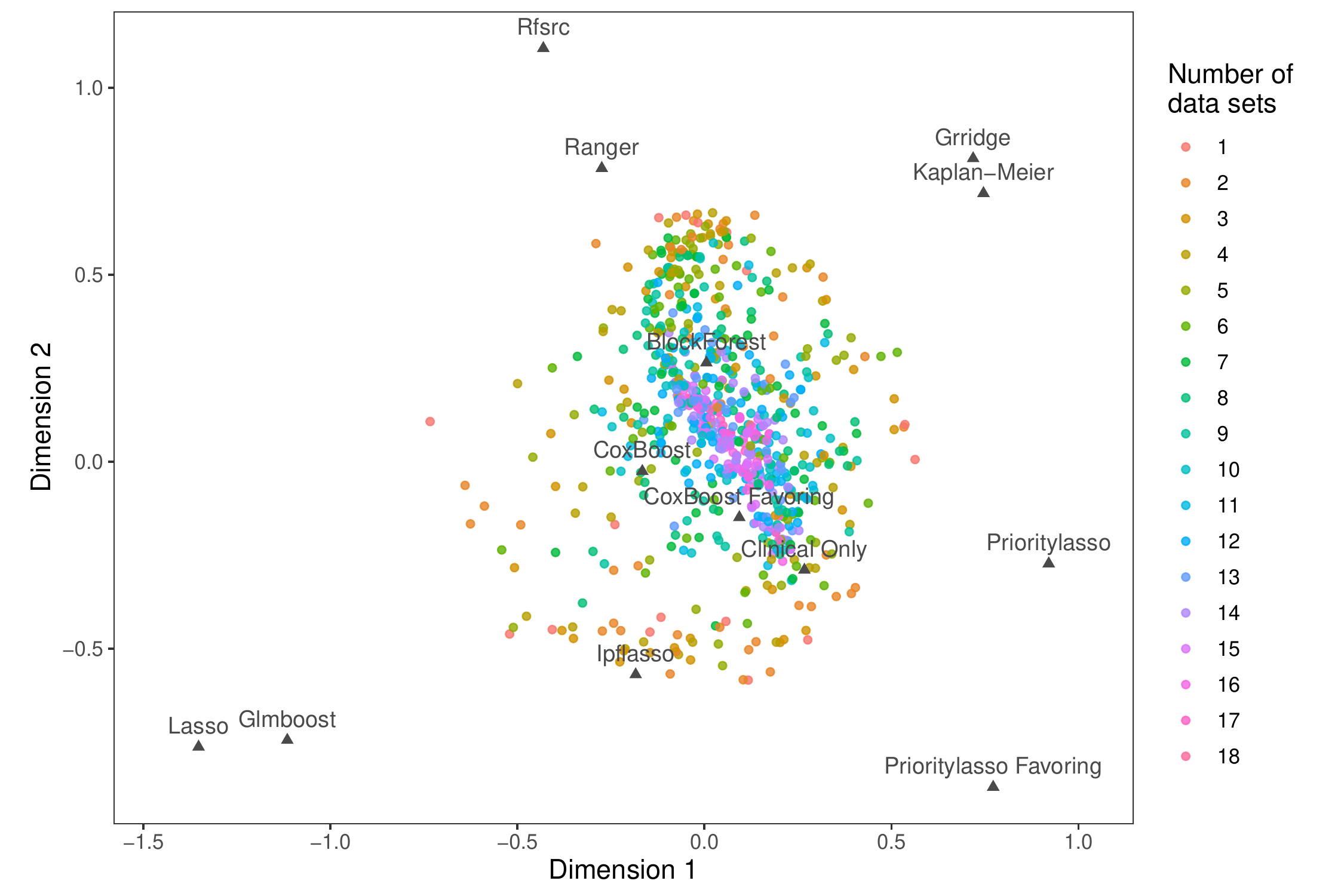}
    \caption{Unfolding solution representing 774 rankings (circles) of 13 methods (triangles) generated by randomly sampling different groups of data sets while performance measure, imputation method, and aggregation method are fixed to their respective default option.}
    \label{fig:unfolding_sampling}
\end{figure}

To gain additional insight concerning the impact of each design and analysis choice, the method rankings are analysed using multidimensional unfolding. Figure~\ref{fig:unfolding} displays the resulting unfolding solution that represents the rankings of all 288 combinations regarding the 13 methods. Before looking at the different colourings of the ideal points in Figure~\ref{fig:unfolding_performance}-\ref{fig:unfolding_aggregation}, we can make some general observations on how the combinations and methods are scaled in the plot (which is identical for each figure). First, the unfolding solution clearly shows that the method rankings can differ widely depending on which combination of design and analysis options is considered, which is consistent with the results presented in Section~\ref{04_results1}. Second, similar to the rank distribution in Figure~\ref{fig:rankdistr}, the unfolding solution indicates that some methods tend to achieve smaller ranks than other methods. This applies specifically to  \Clinicalonly, \CoxBoostfavoring, and \blockForest, which are scaled close to the origin and thus have a small distance to most ideal points. In contrast, other methods such as \Lasso\ and \KaplanMeier\ can be found in the periphery of the plot, indicating that they obtain rather high ranks by most combinations. \\
Of course, the degree to which the presented unfolding solution reflects the actual rankings depends on its goodness-of-fit (a perfect fit usually requires as many dimensions as there are methods, i.e. $dim=M=13$). However, following \cite{Mair2016}, the unfolding solution in Figure~\ref{fig:unfolding} fits the ranking data reasonably well (see the supplementary material for diagnostic figures and measures).\\
An important feature of the unfolding solution in Figure~\ref{fig:unfolding} is that not only the distances between ideal and object points can be interpreted, but also the distances within ideal and object points. This means that, in contrast to the rank distribution in Figure~\ref{fig:rankdistr}, the unfolding solution also provides information about which methods are ranked similarly and which combinations of design and analysis options yield similar rankings. 
We make use of the latter (i.e. the fact that the unfolding solution indicates which combinations yield similar rankings) to assess the impact of each design and analysis choice on the method rankings.
For this purpose, the unfolding solution is supplemented with additional information, which results in Figure~\ref{fig:unfolding_performance}-\ref{fig:unfolding_aggregation}: For each choice, the ideal points are coloured according to the option that was used in the respective combination, with the default option (i.e. the option used in \citealt{Herrmann2020}) coloured in grey. 
Moreover, we connect each ideal point representing the default option to the ideal points representing the alternative options given that the other three choices remain the same. Although this makes the representation dependent on which option is used as the default, for reasons of clarity, we refrain from additionally connecting the alternative options with each other. \\
The resulting plot for the choice of performance measure is displayed in Figure~\ref{fig:unfolding_performance}. The grey lines indicate that the distances between most ideal points corresponding to pairs of ibrier and cindex within one specific setting (i.e.  combinations where the other three choices remain the same) are large. Accordingly, the choice of performance measure strongly impacts the resulting method ranking for most settings. 
Figure~\ref{fig:unfolding_performance} also reveals that the ideal points corresponding to ibrier and cindex form two clearly separated clusters. 
Accordingly, the variability in the method rankings is reduced if the performance measure is fixed. This applies in particular to the cindex, whose corresponding ideal points show considerably less variation than the ideal points corresponding to the ibrier. With regard to the remaining three choices (data sets, imputation method, and aggregation method), this means that their impact is smaller if the cindex is used as performance measure. 
This finding might be explained by the fact that the cindex only measures discriminatory power (see Section~\ref{02_design_analysis_choice}) and might thus be more robust to changes in the remaining design and analysis choices than the ibrier. \\
As can be seen from Figure~\ref{fig:unfolding_datasets}, another important choice that accounts for a large part of the variability in the method rankings is the selection of data sets, especially if the ibrier is used as performance measure (compare with Figure~\ref{fig:unfolding_performance}). 
Figure~\ref{fig:unfolding_datasets} also reveals that within the two clusters corresponding to cindex and ibrier, the ideal points are roughly clustered according to the group of data sets that was used in the respective combination. 
This indicates that keeping the data sets fixed in addition to the performance measure again reduces the variability in the method rankings.
Regarding the type of data sets used in each combination, Figure~\ref{fig:unfolding_datasets} shows that within both clusters of performance measure, the ideal points corresponding to small and large values of each data set characteristic lie approximately opposite to each other while the ideal points representing all 18 data sets are located between them. 
With regard to the choice of data sets, the largest discrepancy between two rankings can thus be expected when comparing the results of two groups that correspond to small and large values of one of the considered data set characteristics. Using all 18 data sets, on the other hand, results in a compromise between the two extremes.\\
As already stated above, the variability in the method rankings is considerably reduced if performance measure and data sets are fixed, which in turn means that the variations caused by using different imputation or aggregation methods are expected to be small. This finding is confirmed by Figure~\ref{fig:unfolding_imputation} and Figure~\ref{fig:unfolding_aggregation}. 
The grey lines indicate that variations in the method rankings caused by deviations from the default imputation or aggregation method mainly arise for ibrier as the performance measure and all groups of data sets except those with many clinical variables or large values of $n$ or $n_e$ (compare with Figure~\ref{fig:unfolding_performance} and Figure~\ref{fig:unfolding_datasets}). In some of the other settings, the impact of the choice of imputation and aggregation method is so small that the ideal points corresponding to different imputation/aggregation methods have the same coordinates (i.e. yield the same ranking). 
This applies in particular to the choice of imputation method, which generally has less impact on the method rankings than the choice of aggregation method, as can be seen from comparing Figure~\ref{fig:unfolding_imputation} and Figure~\ref{fig:unfolding_aggregation}.
\\ 
The distances between ideal points of default and alternative options that are represented as grey lines in Figure~\ref{fig:unfolding_performance}-\ref{fig:unfolding_aggregation} can also be summarised as boxplots, which are displayed in Figure~\ref{fig:unfolding_dist}. This representation provides information that is technically also included in Figure~\ref{fig:unfolding_performance}-\ref{fig:unfolding_aggregation}, but is presented more clearly in Figure~\ref{fig:unfolding_dist}.  For example, it shows for each choice which alternative option used instead of the default option tends to yield the highest variations in the method rankings (e.g. for the choice of imputation method, it is the option that uses the mean of the non-failed iterations as imputation value). Moreover, Figure~\ref{fig:unfolding_dist} reveals that according to the unfolding solution, the largest discrepancy between two rankings generated by only varying one design or analysis option is achieved by using the median instead of the mean as aggregation method. This is an unexpected finding since it has already been stated above and can also be seen from Figure~\ref{fig:unfolding_dist} that in most settings (i.e.  combinations where the other three choices remain the same),  the choice of aggregation method tends to have a smaller impact on the method rankings than the choice of performance measure and data sets. A major drawback of Figure~\ref{fig:unfolding_dist} is that in contrast to Figure~\ref{fig:unfolding_performance}-\ref{fig:unfolding_aggregation}, it does not provide any information about how similar the rankings generated by the alternative options are, nor about how the ranks of the individual methods change.
\linebreakcust
 Of course, all findings concerning the impact of the individual design and analysis choices depend on the number and type of options considered for each choice. Specifically, for the choice of data sets, we only consider a small subset of possible options and we focus, in addition to the 18 original data sets, on groups of approximately equal size (8 to 10 data sets) generated by specific data set characteristics.  We thus complement our analysis by illustrating the impact of the choice of data sets if more options are considered, especially with regard to the number of data sets. For this purpose, we keep performance measure, imputation method, and aggregation method fixed to their respective default option and randomly draw 50 permutations of the 18 original data sets. For each of these permutations we store the method rankings generated by only considering the first $l$ data sets with $l = 1,..,17$, and remove duplicate groups of data sets (which mainly occur for groups with 1, 2, or 17 data sets). This results in 774 rankings including one ranking generated by the 18 original data sets, which are all represented in the unfolding solution in Figure~\ref{fig:unfolding_sampling}. The widely distributed ideal points clearly indicate that the choice of data sets is even more essential if the number of data sets is not restricted and the groups of data sets are not defined based on specific data set characteristics (as it was the case above in Figure~\ref{fig:unfolding}).
As one might have expected, we also observe that the variability in the method rankings increases if the number of data sets decreases. Accordingly, the most extreme rankings (i.e. rankings that differ the most from the ranking generated using all 18 data sets) occur for groups with only a few data sets. Since Figure~\ref{fig:unfolding_performance} revealed that the impact of the choice of data sets strongly depends on the choice of performance measure, we repeat the analysis using cindex as performance measure (see Figure~\ref{fig:unfolding_sampling_cindex} in the supplementary material). Similar to Figure~\ref{fig:unfolding_datasets}, the impact of the choice of data sets is considerably reduced. However, as in Figure~\ref{fig:unfolding_sampling}, the variability in the method rankings increases with decreasing number of data sets.

\section{Discussion}\label{05_discussion}
\subsection{Summary}
In this paper, we addressed the multiplicity of design and analysis options in the context of benchmark studies and the associated risk of over-optimistic results. 
As a preliminary step, we reviewed literature related to the choice of four design and analysis choices that researchers are usually faced with when conducting a benchmark study based on real data sets, namely the choice of data sets, the choice of quantitative performance measure, the choice of imputation method for missing performance values and the choice of aggregation method to generate an overall method ranking.\\
We then used the benchmark study by \cite{Herrmann2020} to illustrate how variable the resulting method rankings of a benchmark study can be when all possible combinations of a range of design and analysis options are considered.
In fact, in this example, the results were so variable that any method could achieve almost any rank, i.e. each method could almost be presented as best or worst method for at least one combination of design and analysis options. 
For the more realistic scenario where the design and analysis options are not systematically examined for each combination but selected in a step-wise optimisation process, we observed that the variability in the method rankings is smaller but still remarkable. \\
In addition to examining the overall variability in the method rankings, we also investigated the individual impact of each choice on the results using multidimensional unfolding.
As might be expected, the choice of performance measure and data sets accounts for a large part of the variability in the method rankings. The impact of the choice of imputation and aggregation method, on the other hand, tends to be considerably smaller but still non-negligible in many settings.
In general, the impact of each choice depends on the options used for the other three choices, with the choice of performance measure affecting the impact of the remaining choices most strongly.
In an additional analysis, we increased the number of considered options for the choice of data sets, which clearly showed that the variability in the method rankings increases if the number of data sets decreases and once again emphasised the importance of the choice of data sets. 

\subsection{Limitations}
Of course, the specific results obtained for the example study by \cite{Herrmann2020} should only be seen as an illustration that cannot be generalised to other benchmark studies. 
Moreover, one possible reason why the method rankings are so variable is that in our example benchmark study, many performance differences are small and the performance values differ widely across data sets, as discussed in the original study by  \cite{Herrmann2020}.
The focus of our study was on ranks, which do not reflect the size of the differences between the methods' performances or the heterogeneity across data sets.
On the one hand, taking these aspects 
into account rather than focusing on ranks may lead to much less variable results, particularly if one relies on statistical tests. On the other hand, the multiplicity of possible analysis options is not limited to the analysis of ranks: there are also plenty of possibly ways to analyse performance differences and the heterogeneity across data sets, even if statistical tests are performed (e.g. paired t-test or Wilcoxon signed-rank test with or without correction for multiple testing, or global tests such as the Friedman test).

\subsection{Negative consequences and possible solutions}
Despite these limitations, our illustration suggests that, as a consequence of the multiplicity of design and analysis options, the results of benchmark studies could be much more variable than many researchers realise. Combined with questionable research practices (e.g. the selective reporting of results or the targeted modification of specific design and analysis components), this potentially high variability of benchmark results can lead to biased interpretations and over-optimistic conclusions regarding the performance of some of the considered methods. Given the high level of evidence that is attributed to neutral benchmark studies \citep{Boulesteix2017}, a ``neutral'' benchmark study that is in fact biased could thus negatively affect both methodological and applied research by misleading method users and developers \citep{Weber2019}. \linebreakcust
Fortunately, there are several strategies to prevent over-optimistic benchmark results that arise from the multiplicity of design and analysis options, some of which are already applied by many researchers, including \cite{Herrmann2020}. For example, strategies inspired from blinding in clinical trials can help to reduce non-neutrality and/or the potential to exploit the multiplicity of possible options. Specifically, blinding could be realised by labelling the methods with non-informative names (e.g. Method A, Method B, etc.) such that researchers have no information about the performance of each method until the end of the study \citep{Boulesteix2017}. If the benchmark study is based on simulated data, researchers could also be blinded to the data generation process, which prohibits the possibility to tune the parameters of selected methods according to the known ground truth \citep[e.g.][]{Kreutz2020}. \\
The remaining strategies to prevent over-optimistic results can be summarised using the work of \cite{Hoffmann2021}, who formalise the effect of both random sources of uncertainty (including sampling uncertainty) and epistemic sources of uncertainty (resulting in a  multiplicity of possible analysis strategies and thus opening the door to questionable research practices) on the replicability of research findings. 
They outline six steps researchers from all empirical research fields  can take to make their own research more replicable and credible. 
In brief, researchers should (1)  be aware of the multiplicity of possible analysis strategies, (2) reduce uncertainty, (3) integrate uncertainty, (4) report uncertainty, (5) acknowledge uncertainty, and (6) publish all research code, data and material. 
Although \cite{Hoffmann2021} focus on applied rather than methodological research,
we argue that their recommended steps can also be applied to address the sources of uncertainty that arise from the design and analysis of benchmark studies. 
\paragraph*{Step 1} 
In the context of benchmark studies, the first step to reduce the risk of over-optimistic results is to simply be aware of the multiplicity of possible design and analysis options and the potential for questionable research practices.
We can only speculate about how much awareness for this issue is already present in methodological research but hope that this paper contributes to raising it. 
\paragraph*{Step 2}
The second step suggested by \cite{Hoffmann2021} is to reduce sources of uncertainty. 
In the context of benchmark studies, this could be realised by consulting existing benchmarking guidelines found in literature. However, as discussed in this paper, guidelines for many issues relevant in practice are still lacking. We claim that more guidance and standardised approaches are needed in this context.    
Regarding the choice of data sets, uncertainty could be reduced if the number of data sets to include in the study would be consequently based on statistical considerations such as power calculation \citep[e.g.][]{Boulesteix2015a} and if data sets would be selected according to strict and well-considered inclusion criteria. 
Both aspects are facilitated if structured and well-documented databases exist for the type of data to be studied. 
\paragraph*{Step 3} 
As a third step, \cite{Hoffmann2021} recommend to integrate remaining sources of uncertainty that could not be reduced in the second step. 
Analysis approaches such as confidence intervals, statistical tests or boxplots that take the heterogeneity of performance values across data sets into account can be seen as first steps towards integrating the uncertainty regarding the choice of data sets.  
However, they do not provide much information about how the benchmark results would change if only certain subgroups of data sets would be considered.
A more advanced but less common way to integrate uncertainty regarding the choice of data sets is to analyse the relationship between method performance and data set characteristics \citep[e.g.][]{Oreski2017, Kreutz2020,Eugster2014}. 
Concerning the choice of evaluation criteria (including quantitative performance measures), the aggregation of method rankings resulting from different criteria into an overall ranking can be seen as an attempt towards integrating uncertainty. 
However, to our knowledge, currently existing approaches such as consensus rankings \citep{Hornik2007} do not provide any measure of uncertainty.  
\paragraph*{Step 4} 
For all sources that cannot be adequately 
integrated, \cite{Hoffmann2021} suggest to systematically report the results of alternative analysis strategies, which, in the context of benchmark studies, would be alternative design and analysis options. 
While reporting the results of alternative analysis strategies, e.g. in the form of a sensitivity analysis, is a common procedure in applied research \citep{Hoffmann2021}, to our knowledge it is rarely performed in benchmark studies (especially if they are based on real data sets). 
However, considering the lack of ways to reduce and integrate uncertainty when designing and analysing benchmark studies, adequately reporting the results of alternative options seems to be all the more important.
One reason for the lack of uncertainty reporting in benchmark studies could be that, to our knowledge, no suitable framework has been available so far. 
This gap could be filled by the framework based on multidimensional unfolding that we used in this paper.
It can be seen as a systematic version of standard sensitivity analysis that allows to graphically assess the variability of the method rankings with respect to a large number of different combinations of design and analysis options.
It also provides information about the individual impact of each choice on the method ranking and thus enables researchers to analyse when and how using alternative options for a specific choice affects the results. 
In this way, the risk of misleading readers is reduced and the benchmark results become even more reliable and valuable.
Moreover, using the framework allows to identify critical choices that substantially affect the results and should therefore be particularly well justified in future benchmark studies and be given more consideration in benchmarking guidelines.  
\paragraph*{Step 5} 
The next important step suggested by \cite{Hoffmann2021} is to accept the inherent uncertainty of scientific findings.
In the context of benchmark studies, this implies that researchers should clearly state that the benchmark results are conditional on the selected design and analysis options \citep{Boulesteix2013, Hornik2007}. 
In this vein, researchers should also acknowledge that just as in applied research, generalisations from a single study are usually not appropriate \citep{Amrhein2019,Hoffmann2021}.
This emphasises the need for more high-quality benchmark studies and for meta-analyses of benchmark studies \citep[e.g.][]{Gardner2019}, which, however, are still  rare and unfortunately sometimes not considered as full-fledged research by the scientific community \citep{Boulesteix2020}.
Another aspect also related to the acceptance of uncertainty is to recognise that statistical inference within exploratory  analyses should be treated with great caution \citep{Amrhein2019,Hoffmann2021}. Similar to applied research, strictly confirmatory benchmark studies could be realised by pre-registration of design- and analysis plans, as recently implemented in the context of the so-called pre-registration experiment (see \url{https://preregister.science}) or through the \lq\lq registered report'' publication format \citep{chambers2013registered}, which has meanwhile been adopted by several interdisciplinary journals that also accept computational papers.\\
It is also important to recall that there is usually no best method for all scenarios and data sets (the well-known ''no free lunch" theorem; \citealp{Wolpert2002}).
Especially for data sets and evaluation criteria, it might thus be advisable to accept the uncertainty that is associated with their choice by putting more focus on the analysis of the individual strengths and weaknesses of each method than on an aggregated overall ranking.
This can for example be realised by individually analysing the rankings generated by each evaluation criterion and by investigating the relationship between method performance and data set characteristics (see Step 3).
\paragraph*{Step 6} As a final step, the publication of codes and (if possible) data sets that ideally allow the extension to alternative options and additional methods can reduce the impact of over-optimism since it enables readers to run alternative analyses and to reveal potentially biased results.

\section{Conclusion}\label{06_conclusion}
In conclusion, our illustration suggests that benchmark results can be highly variable with respect to design and analysis choices, which can lead to biased interpretations and over-optimistic conclusions. However, there is a wide range of strategies that can help to avoid these pitfalls. We hope that our proposed framework makes a useful contribution towards this objective. While a certain amount of over-optimism can probably never be completely avoided, addressing this problem will lead to more reliable and valuable benchmark results.

\section*{Funding Information}

{\label{974317}}
This work was supported by the German Federal Ministry of Education and Research (01IS18036A) and by the German Research Foundation (BO3139/4-3, BO3139/7-1, BO3139/6-2) to ALB. The authors of this work take full responsibilities for its content.

\section*{Acknowledgements}

{\label{749861}}

The authors thank Anna Jacob for language correction.

\FloatBarrier
\bibliographystyle{apalike}
\bibliography{bibliography}

\newpage
\section*{Supplementary material}
\beginsupplement
\subsection*{Weighted imputation method for missing cindex performance values }
For the cindex, where 1 corresponds to the best possible value and 0.5 to random prediction, the imputed value for the considered combination of data set and method that corresponds to the proposed ``weighted imputation method'' is
\begin{align}
    x_{impute} = 0.5+ (\frac{\sum_{i\in \mathcal{I}}x_i}{|\mathcal{I}|}-0.5)_+ \cdot(1-r),
\end{align}
where $\mathcal{I}$ is the set of indices of the non-failed iterations,  $x_i$ is the cindex value for iteration $i \in \mathcal{I}$ and $r$ is the proportion of missing values.

\subsection*{Additional figures step-wise optimisation}
\begin{figure}[ht]
	\centering
	\includegraphics[trim= 0cm 0cm 0.2cm 0cm, clip, width=0.8\textwidth]{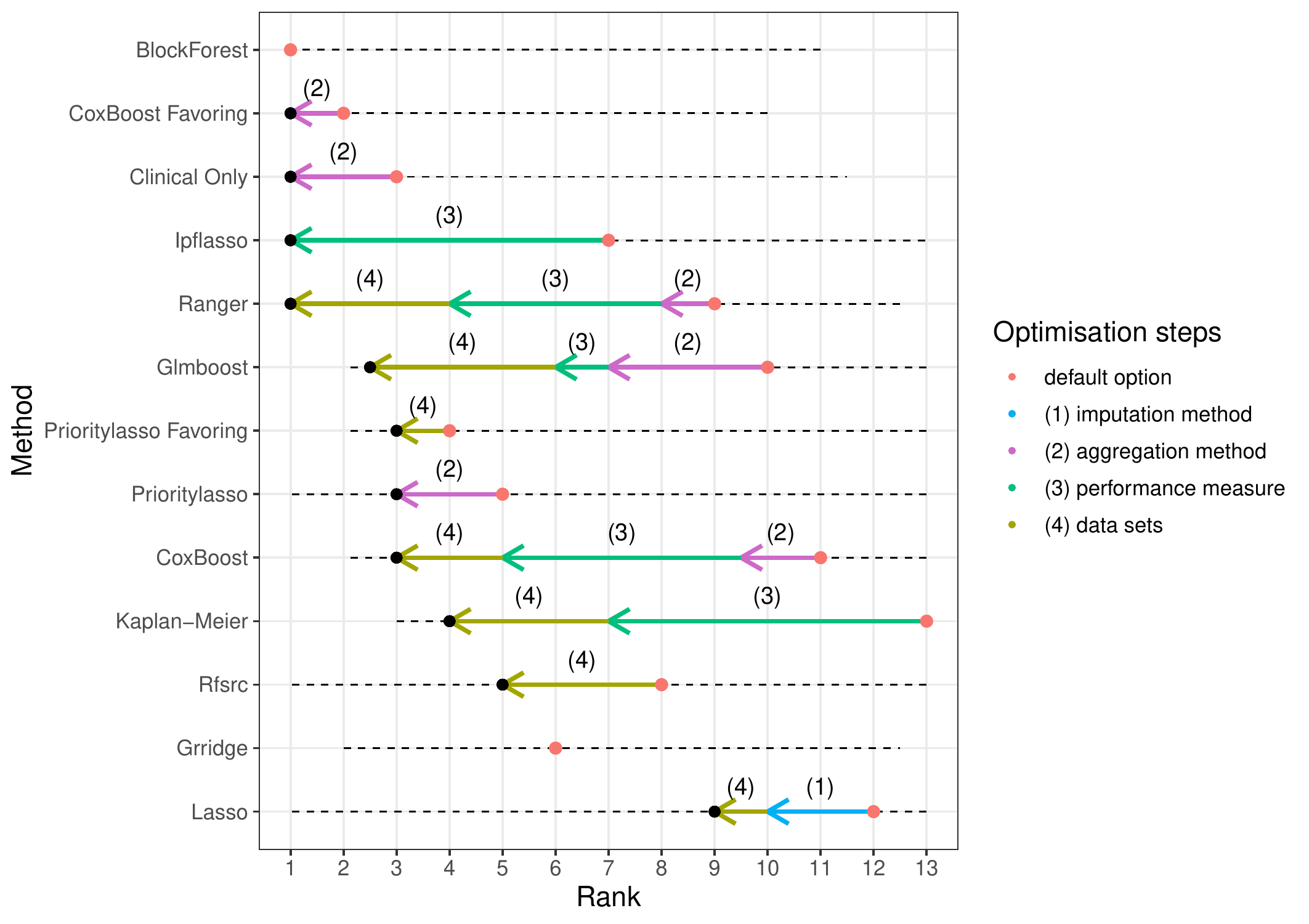}
\caption{Step-wise optimisation of method ranks by (1) imputation method (blue), (2) aggregation method (pink), (3) performance measure (green), and (4) data sets (yellow). The dotted line corresponds to the smallest and highest possible ranks when all 288 combinations are considered. Missing steps indicate that they did not lead to an improved rank. Default options correspond to \protect\cite{Herrmann2020} except performance measure, which is set to cindex.}
			\label{fig:stepwise imp_agg_eval_group_cindex} 
\end{figure}

\begin{figure}[ht]
	\centering
	\includegraphics[trim= 0cm 0cm 0.2cm 0cm, clip, width=0.8\textwidth]{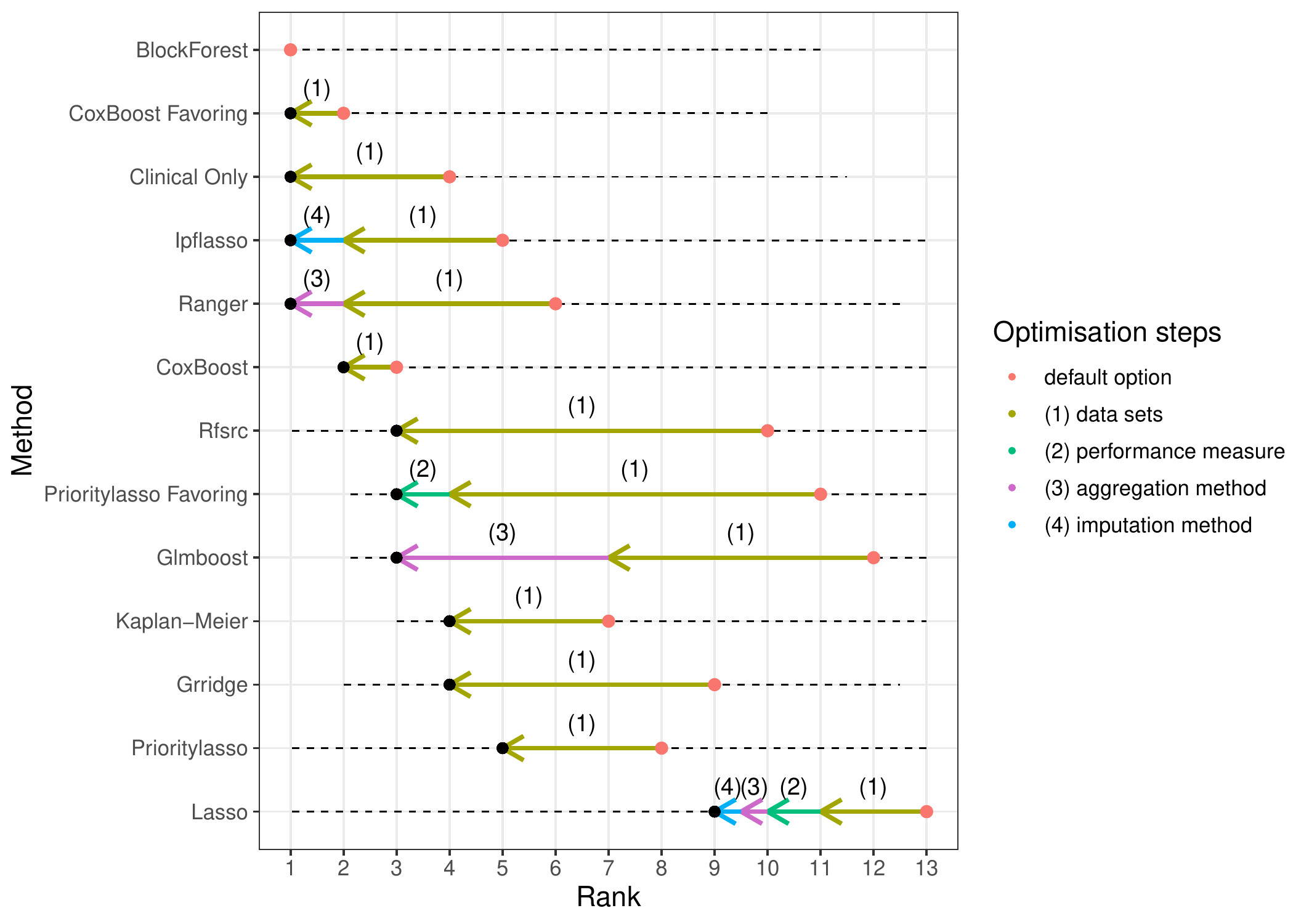}

\caption{Step-wise optimisation of method ranks by (1)  data sets (yellow), (2) performance measure (green), (3)~aggregation method (pink), and (4) imputation method (blue). The dotted line corresponds to the smallest and highest possible ranks when all 288 combinations are considered. Missing steps indicate that they did not lead to an improved rank. Default options correspond to \protect\cite{Herrmann2020}.}
	\label{fig:stepwise group_eval_imp_agg} 
\end{figure}
\clearpage
\subsection*{Additional figures unfolding}

\begin{figure}[ht]
    \centering
    \includegraphics[width = 0.9\textwidth]{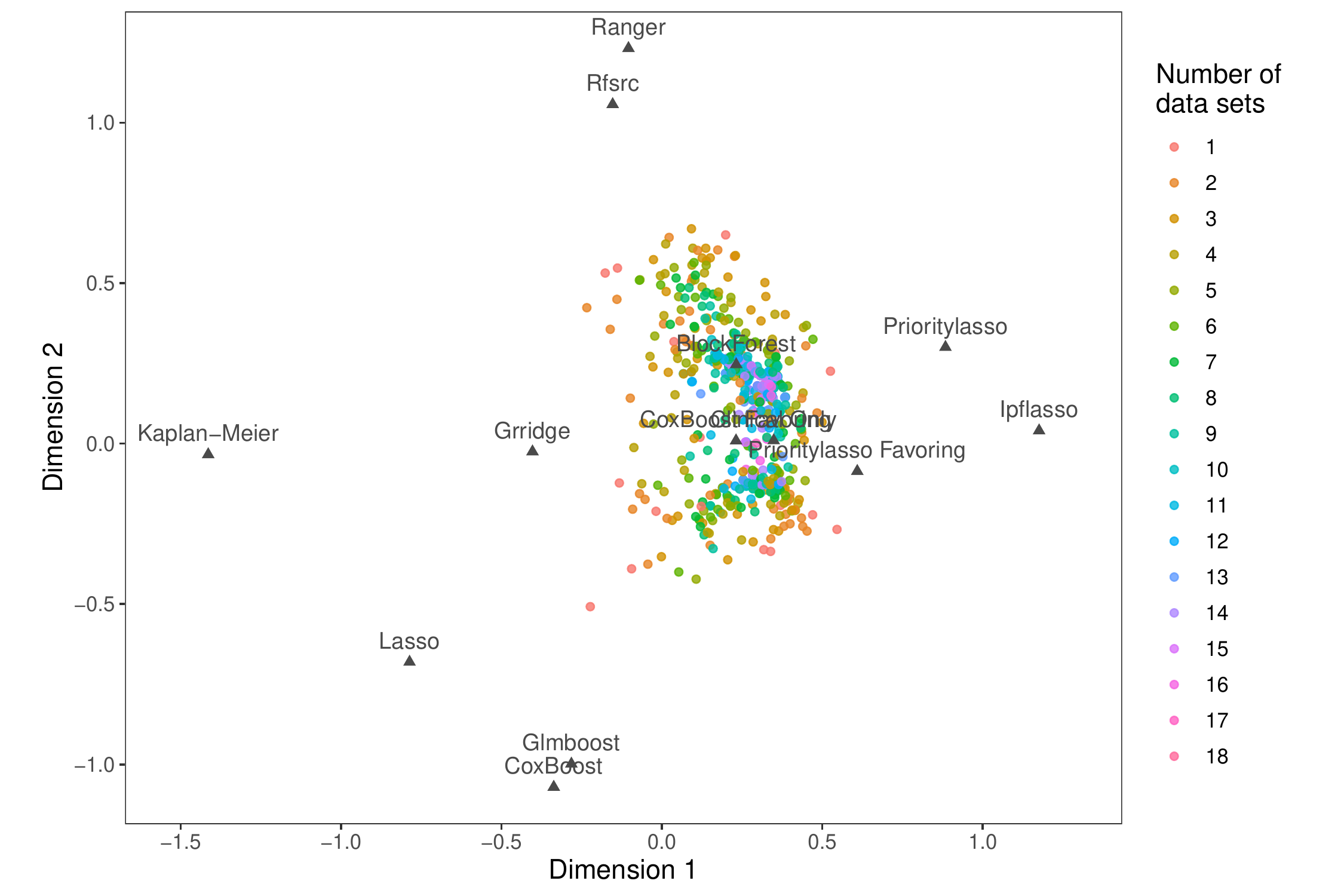}
    \caption{Unfolding solution representing 774 rankings (circles) of 13 methods (triangles) generated by randomly sampling different groups of data sets while imputation method and aggregation method are fixed to their respective default option and performance measure is set to cindex.}
    \label{fig:unfolding_sampling_cindex}
\end{figure}

\subsection*{Goodness-of-fit unfolding solutions}
The assessment of the goodness-of-fit is based on \cite{Mair2016}. We assess the fit of three unfolding models presented in this paper:
\begin{itemize}
\item Model 1: Unfolding solution representing the rankings of 288 combinations of design and analysis
options regarding 13 methods
\item Model 2: Unfolding solution representing 774 rankings of 13 methods generated
by randomly sampling different groups of data sets while performance measure, imputation
method, and aggregation method are fixed to their respective default option
\item Model 3: Unfolding solution representing 774 rankings of 13 methods generated by randomly sampling different groups of data sets while imputation method and aggregation method are fixed to their respective default option and performance measure is set to cindex
\end{itemize}
\paragraph{Permutation test}
We test the null hypothesis that the unfolding solution is obtained from a random permutation of dissimilarities. Rejecting the null hypothesis provides some evidence that the unfolding solution captures a structural signal. For all three unfolding models, the resulting p-value is $< 0.001$.

\begin{figure}[ht]
    \centering
    \includegraphics[width = 0.6\textwidth]{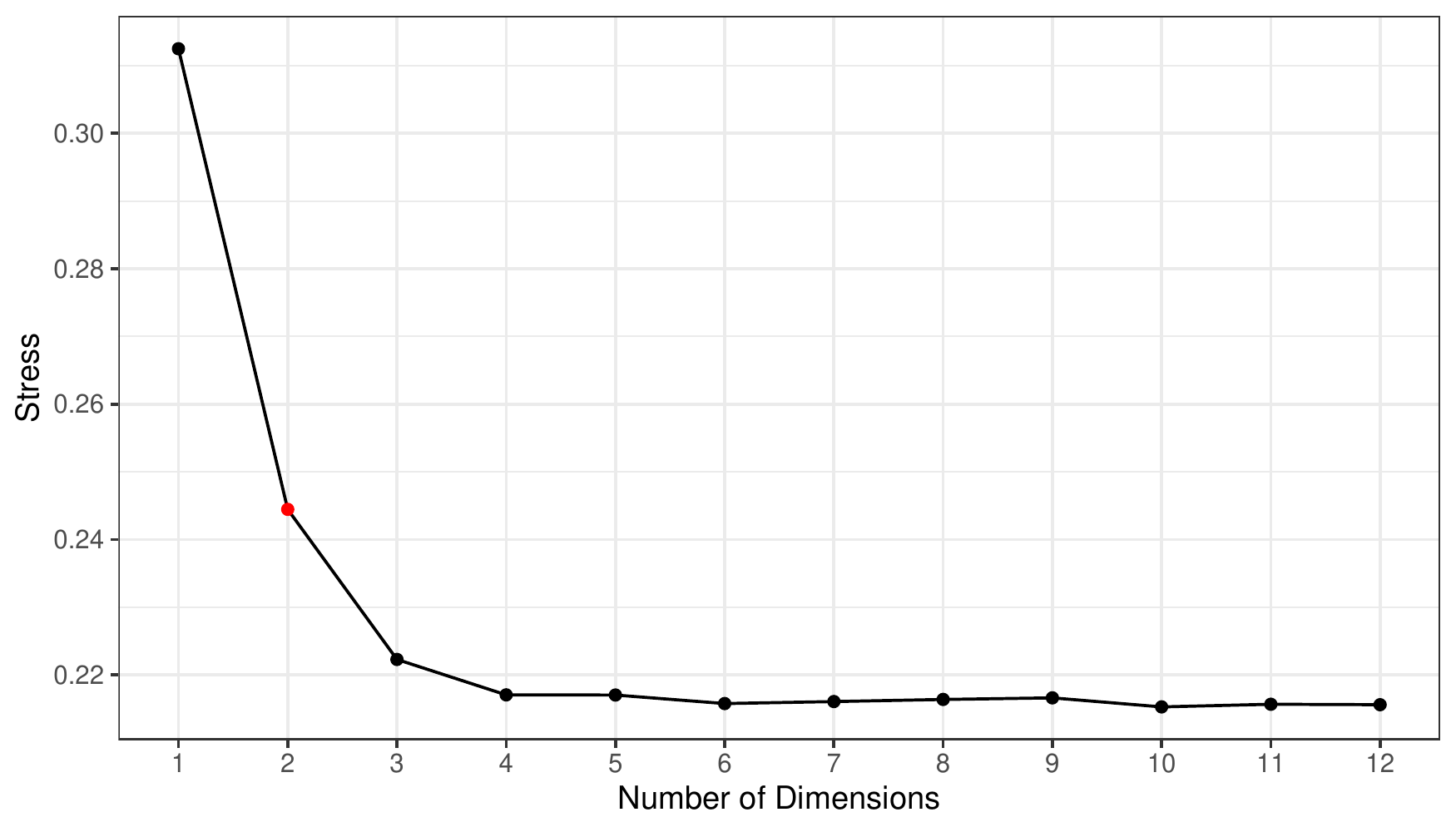}
     \caption{Scree plot for unfolding model 1. The stress value for $dim =2$, which was used in our application, is coloured in red.}
     \label{fig:scree_unfolding}
\end{figure}

\begin{figure}[ht]
    \centering
    \includegraphics[width = 0.6\textwidth]{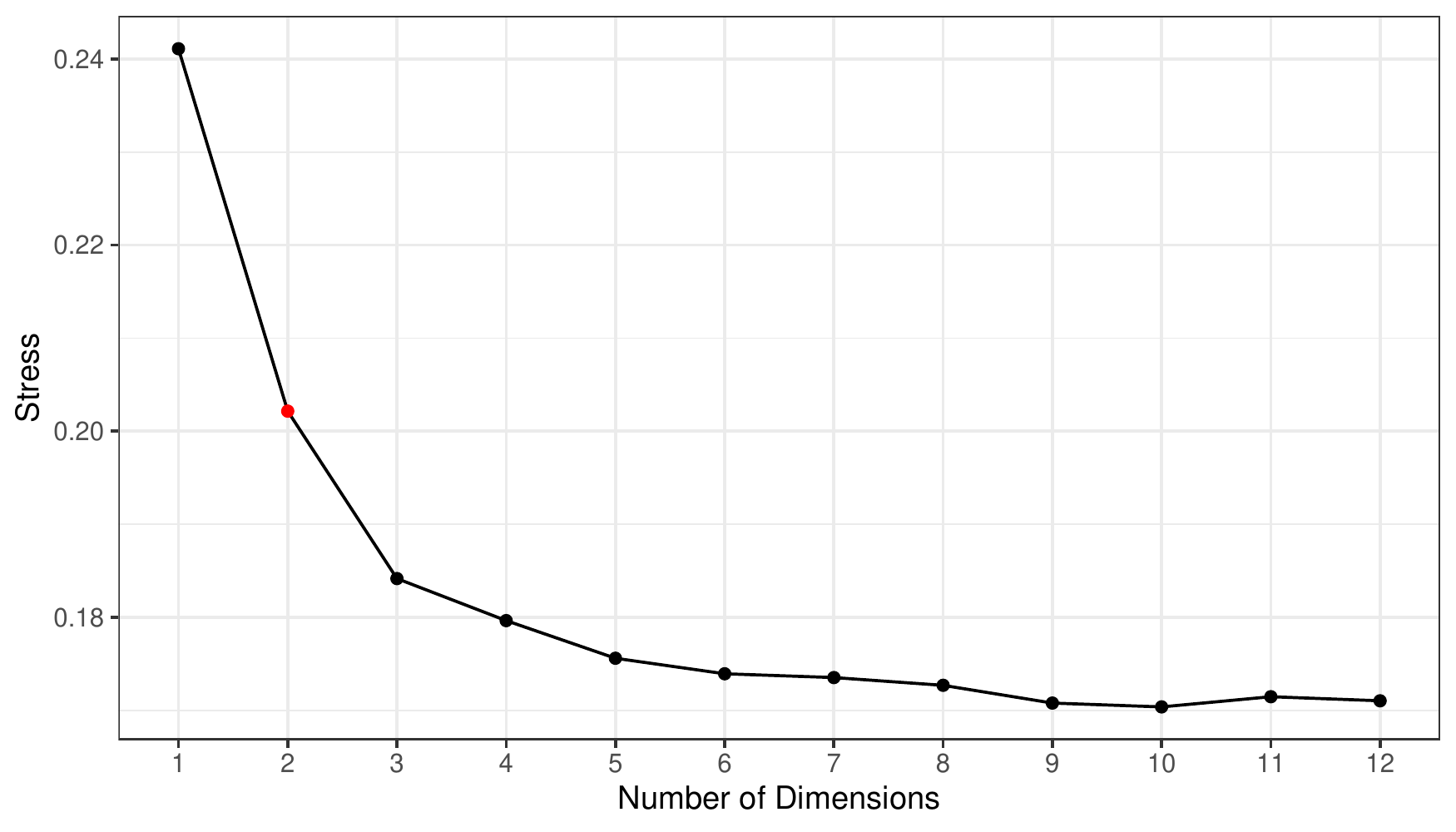}
        \caption{Scree plot for unfolding model 2. The stress value for $dim =2$, which was used in our application, is coloured in red.}
     \label{fig:scree_unfolding_sampling_ibrier}
\end{figure}

\begin{figure}[ht]
    \centering
    \includegraphics[width = 0.6\textwidth]{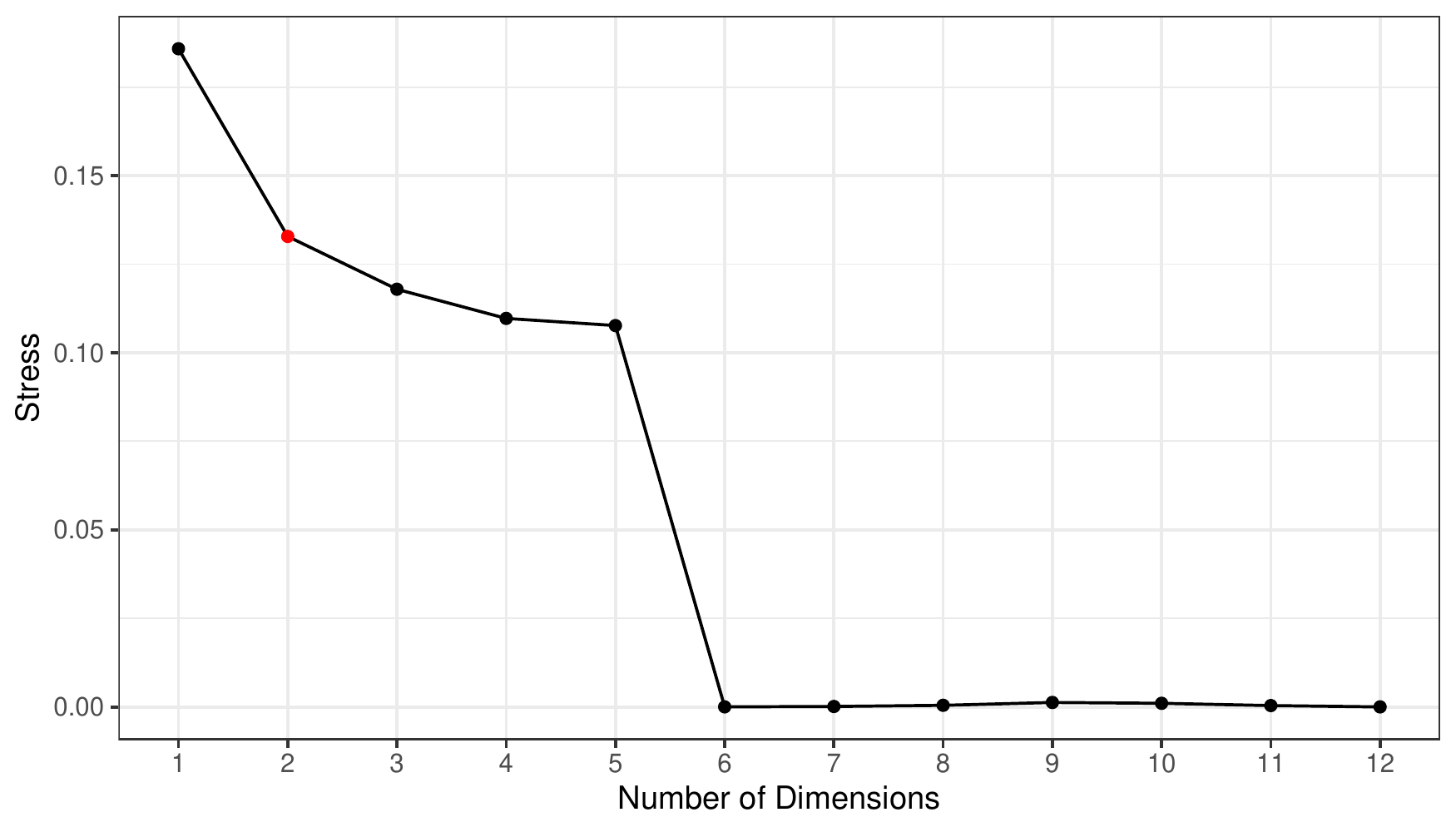}
       \caption{Scree plot for unfolding model 3. The stress value for $dim =2$, which was used in our application, is coloured in red.}  \label{fig:scree_unfolding_sampling_cindex}
\end{figure}

\paragraph{Scree plots} We generate scree plots with varying number of dimensions (i.e. $dim = 1,\dots, 12$, since $dim =13$ results in a stress value of $0$). Ideally, we would see an elbow at $dim = 2$ (the dimension chosen for the unfolding models in this paper), which would indicate that additional dimensions represent only random components of the data \citep{Borg2013}. Although no clear elbow is visible in Figure~\ref{fig:scree_unfolding}-\ref{fig:scree_unfolding_sampling_cindex}, the scree plots indicate that the stress is already considerably low for $dim = 2$. Note that in Figure \ref{fig:scree_unfolding_sampling_cindex}, the iteration limit was reached when running the unfolding models for $dim \geq 6$ and the stress is close to 0, which might indicate degenerate solutions. 

\begin{figure}[ht]
    \centering
    \includegraphics[width = 0.9\textwidth]{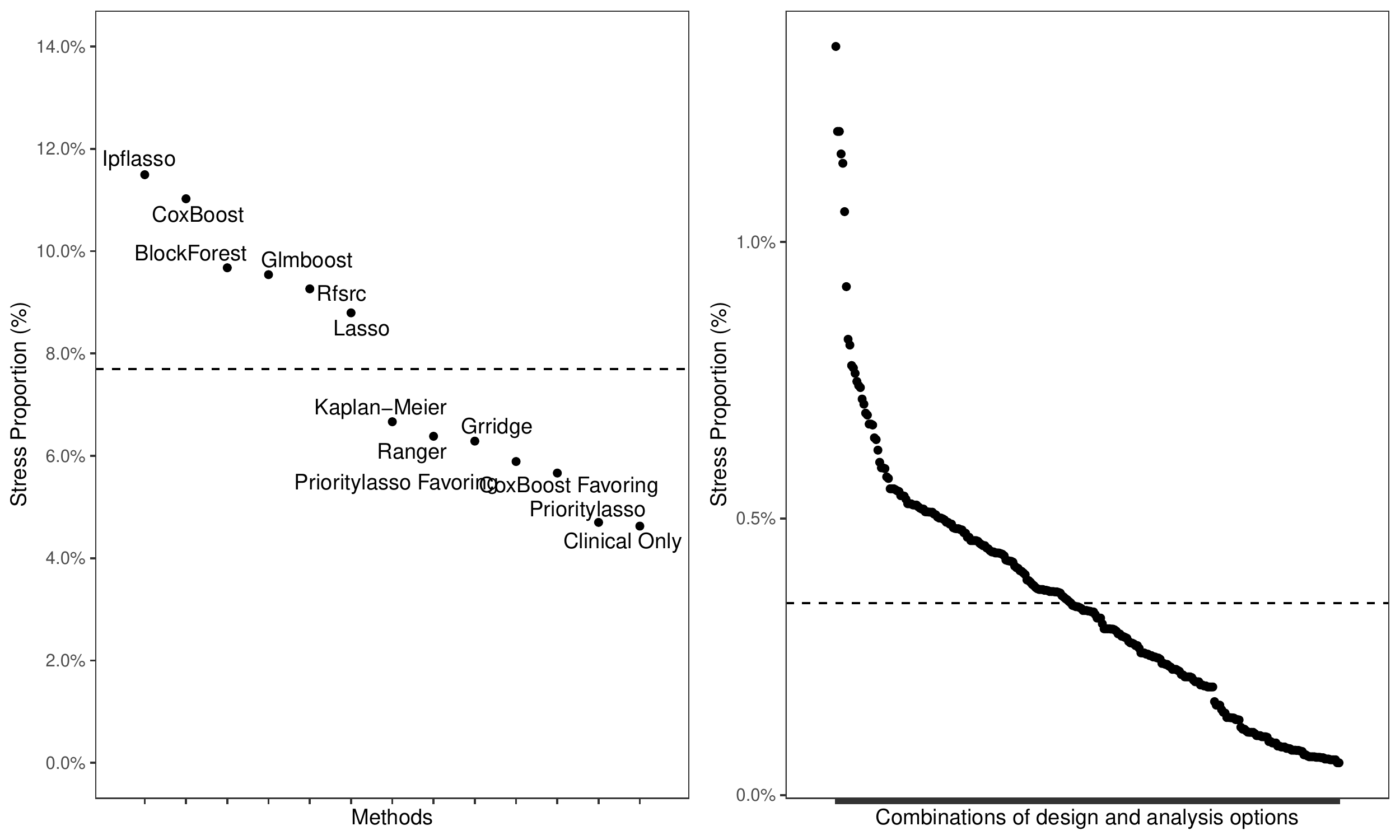}
    \caption{Stress-per-point for methods (left) and combinations of design and analysis options (right) for unfolding model 1. The greater the stress proportion, the more the point contributes to the misfit of the unfolding solution. The dotted line represents the stress proportion if every method/combination contributed equally to the misfit.}
    \label{fig:spp_unfolding}
\end{figure}

\begin{figure}[ht]
    \centering
    \includegraphics[width = 0.9\textwidth]{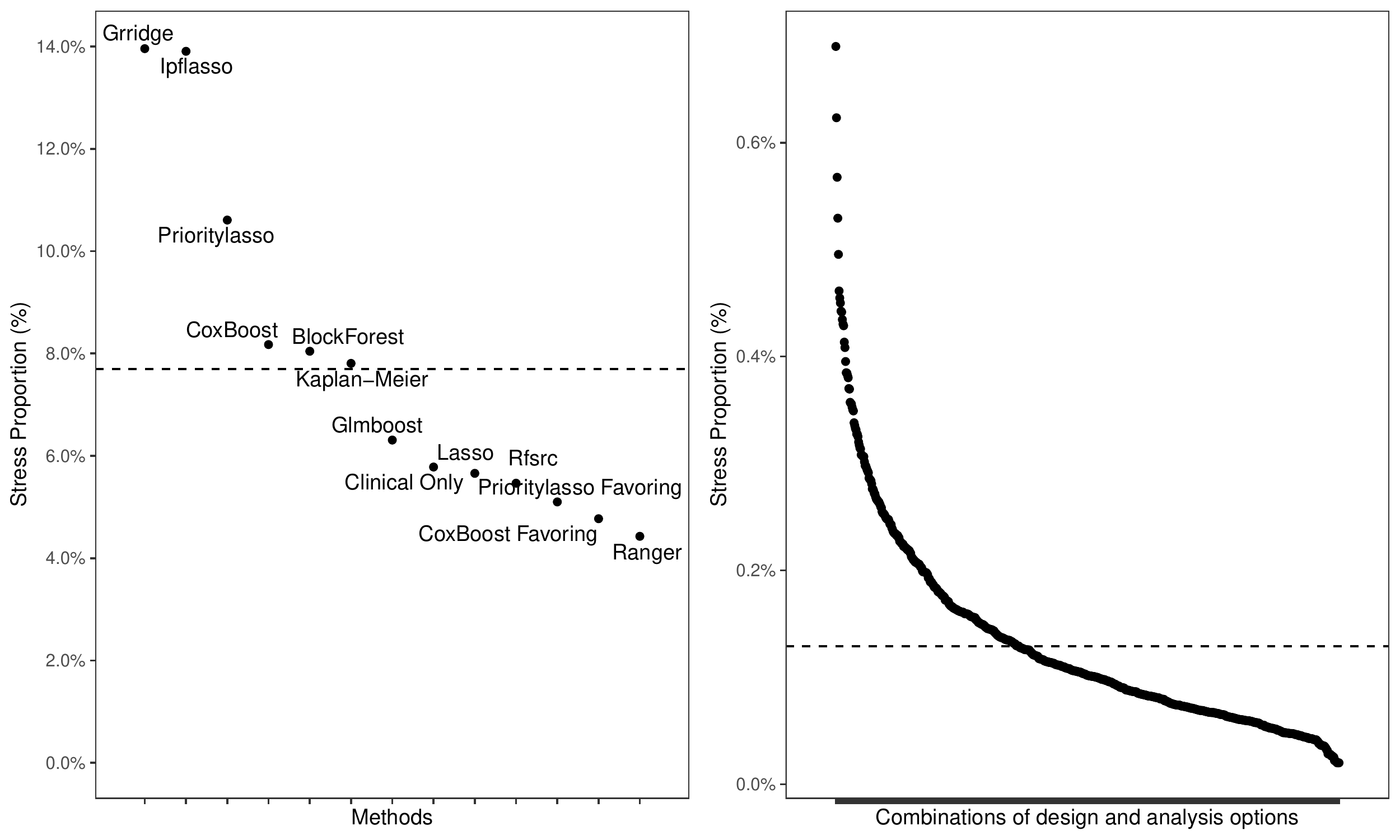}
    \caption{Stress-per-point for methods (left) and combinations of design and analysis options (right) for unfolding model 2. The greater the stress proportion, the more the point contributes to the misfit of the unfolding solution. The dotted line represents the stress proportion if every method/combination contributed equally to the misfit.}
    \label{fig:spp_unfolding_sampling}
\end{figure}

\begin{figure}[ht]
    \centering
        \includegraphics[width = 0.9\textwidth]{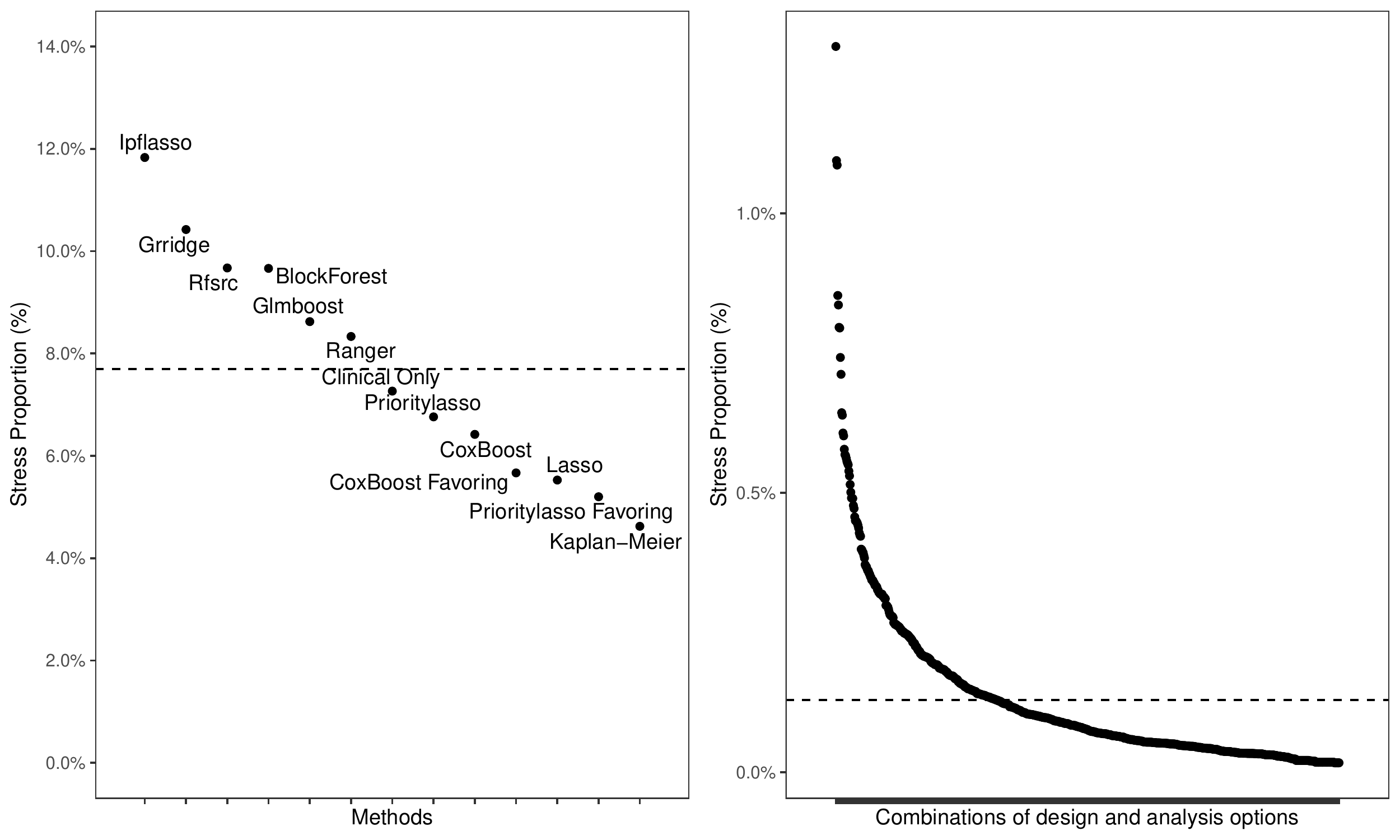}
   \caption{Stress-per-point for methods (left) and combinations of design and analysis options (right) for unfolding model 3. The greater the stress proportion, the more the point contributes to the misfit of the unfolding solution. The dotted line represents the stress proportion if every method/combination contributed equally to the misfit.}
     \label{fig:spp_unfolding_sampling_cindex}
\end{figure}

\paragraph{Stress-per-point}
To check for influential points that should be subject to special consideration, we can look at the stress-per-point values (SPP). The SPP values are assessed separately for subjects (here: combinations of design and analysis options) and objects (here: methods). As can be seen from Figure ~\ref{fig:spp_unfolding}-\ref{fig:spp_unfolding_sampling_cindex}, there are no extreme outliers for any of the three unfolding models presented in this paper. On the method side, all stress proportions are smaller than 14\%, and on the combination side, most stress proportions are smaller than 1\%.
\end{document}